\documentclass[nofootinbib,eqsecnum,tightenlines,superscriptaddress,10.5pt]{revtex4}

\usepackage{hyperref}
\usepackage{graphicx}
\usepackage{amsmath,amssymb,amsfonts,amsthm,stmaryrd,mathtools,bm,physics,tensor}
\usepackage{color}
\usepackage{tcolorbox}
\allowdisplaybreaks[1]

\def\be#1\ee{\begin{align}#1\end{align}}

\def\ba{\begin{eqnarray}}
\def\ea{\end{eqnarray}}
\def\nn{\nonumber}

\def\q{\quad}

\begin{document}

\title{Area-metric gravity revisited}

\author{Johanna N.~Borissova}
\email{jborissova@perimeterinstitute.ca} 
\affiliation{Perimeter Institute, 31 Caroline Street North, Waterloo, ON, N2L 2Y5, Canada}
\affiliation{Department of Physics and Astronomy, University of Waterloo, 200 University Avenue West, Waterloo, ON, N2L 3G1, Canada}
\author{Bianca Dittrich}
\email{bdittrich@perimeterinstitute.ca}
\affiliation{Perimeter Institute, 31 Caroline Street North, Waterloo, ON, N2L 2Y5, Canada}
\author{Kirill Krasnov}
\email{Kirill.Krasnov@nottingham.ac.uk}
\affiliation{School of Mathematical Sciences, University of Nottingham, NG7 2RD, UK}

\begin{abstract}

Area metrics are an intriguing generalization of length metrics which appear in several quantum-gravity approaches. We describe the space of diffeomorphism-invariant area-metric actions quadratic in fluctuations and derivatives.  A general theory is found to be specified by four parameters, two of which are mass parameters for the non-length degrees of freedom. We find that a two-parameter subclass of theories exhibits an additional ``shift" symmetry of the kinetic term, and leads to a ghost-free graviton propagator for the effective theory obtained after integrating out the non-length degrees of freedom. One of the two parameters determines the strength of parity violations, the other defines a mass parameter for the non-length degrees of freedom. The same type of action has been found to appear from modified Plebanski theory and in the continuum limit of (effective) spin foams. In this case the two parameters can be identified with the Barbero-Immirzi parameter and a combination of the Barbero-Immirzi parameter and the Planck mass, respectively.

We moreover find that area-metric actions in Lorentzian (but not in Euclidean) signature feature wrong-sign kinetic and mass terms for the non-length degrees of freedom.  Nevertheless, despite a coupling of these degrees of freedom to the length metric, the linearized dynamics turns out to be stable for the above subclass of actions.

\end{abstract}

\maketitle

\vspace{-0.5cm}

\tableofcontents

\section{Introduction}\label{Sec:Introduction}

 General relativity is built on the concept of a length metric. Many approaches to quantum gravity suggest however generalizations of the space of length metric geometries. One generalization which appears across a number of approaches are area metrics.  Similar to the length metric measuring the length of tangent vectors and angles between these, the area metric measures the areas of parallelograms in tangent space, and dihedral angles between such parallelograms.
 
 In four dimensions, cyclic~\footnote{See below for a definition of cyclic. In this paper we restrict to cyclic area metrics, a general area metric has 21 components.} area metrics have 20 components, as compared to the 10 components of the length metric. Each length metric induces an area metric, in this sense area metrics constitute (in four spacetime dimensions) a proper generalization of the length-metric space.

Area metrics have been proposed to describe the phenomenological effects of quantum gravity in e.g.~\cite{Schuller}. They also appear in string theory~\cite{Schuller} and holography, where they are essential for the reconstruction of geometry from entanglement~\cite{Ryu:2006bv}. Area variables serve also as the basic variables in loop quantum gravity~\cite{AreaDiscrete,Ryan1,Padua23} and spin foams~\cite{PerezReview,AreaAngle,EffSF1}.

The spin-foam path integral sums over loop quantum gravity data associated to simplices in a given triangulation. These simplex data define an area metric for each simplex, that is an area metric at the microscopic level \cite{Padua23}. But the area metric also appears at the macroscopic level:
 \cite{AR1,AR2} revealed that the continuum limit of the Area-Regge action~\cite{BarretEtAl, AsanteAreaRegge}, which describes the semi-classical regime of spin foams, gives rise to an action for an area metric. This area metric action leaves only the length metric degrees of freedom massless, whereas the remaining area metric degrees of freedom obtain a (Planck-scale) mass. One thus recovers the Einstein-Hilbert action, but also (Planck-scale suppressed) corrections, quadratic in the Weyl curvature tensor~\cite{AR2}.

Spin-foam inspired area metric dynamics can also be directly derived from the continuum. The Plebanski  action~\cite{Plebanski} of general relativity underlies spin foam dynamics. The Plebanski formalism  turns a topological action into an action for general relativity by imposing so-called simplicity constraints, which reduce the configuration space to the space of length metrics (or rather tetrad variables). Modified Plebanski theories~\cite{Krasnov1} replace these simplicity constraints by potential terms added to the action. Thus one enlarges the configuration space of length metrics, but equips all additional degrees of freedom with mass via the added potential terms. 

The work~\cite{BorDitt} chooses a splitting of the simplicity constraints into two parts, and imposes one part sharply and the other part via potential terms. The imposition of the first part leads to a configuration space of area metrics and thus an action in terms of area metrics. The imposition of the second part via potential terms adds mass to all the degrees of freedom in addition to the length metric. Integrating out these additional degrees of freedom  allows to find the perturbative effective action for the length metric and one finds again the Einstein-Hilbert action plus a Weyl squared term, suppressed by the mass and inverse derivatives. Interestingly, this combination is such that it leads to a ghost-free linearized theory, where only the graviton degrees of freedom are propagating~\cite{BorDitt}. The same type of effective action was found earlier for modified chiral Plebanski theory \cite{FreidelMod,Krasnov2}.

The enlargement of the length-metric configuration space to area metrics can be understood as a consequence of a fundamental quantum uncertainty.  In fact, the quantization of the simplicity constraints leads to a partially second-class constraint algebra with an anomaly ~\cite{Ryan2,Ryan3} controlled by the so-called Barbero-Immirzi parameter~\cite{BIParameter}. This allows for imposing the second-class constraints only weakly, which leads to an enlarged configuration space over which the spin-foam path integral is defined. An alternative argument for this enlargement starts from the discrete (and asymptotically equally spaced) spectrum for the area operators in loop quantum gravity~\cite{EffSF1}. Imposing all the simplicity constraints sharply would lead to diophantine equations for the discrete eigenvalues of the area operators. The resulting solution space is too small to support semi-classical states, one is thus forced to consider an enlargement of the configuration space.

Having motivated the appearance of area metrics in quantum gravity, and in particular in loop quantum gravity, we now ask what type of actions one can construct using area metrics.  This question has already been addressed in the framework of constructive gravity \cite{SchullerProc}. Thus, 
general area metric actions, up to second order in fluctuations and derivatives have been constructed both in the canonical \cite{ConGrav1} and covariant \cite{ConGrav2} version. We revisit the same problem. We show that there is a simple solution to the problem, based on only the representation theory of the Lorentz group, together with the requirement of the diffeomorphism invariance. The quadratic action we describe contains only a handful of parameters, as compared to the 37 parameters in \cite{ConGrav2}.

Thus, in this work general covariance is used as the main guiding principle.  Indeed it is well known that imposing diffeomorphism invariance on actions of length metrics significantly restricts their possible form. According to Lovelock's theorem~\cite{Lovelock1971} 
only the Einstein field equations can arise as Euler-Lagrange equations from a local, second-order in derivatives Lagrangian for the metric tensor in four spacetime dimensions. More straightforwardly, as is well-known, at quadratic order one can completely fix the form of the action for length metric perturbations using diffeomorphism invariance. To that end, consider the length metric tensor expanded around a flat background~\footnote{Here we have choosen Euclidean signature and expanded the metric around the flat Euclidean metric $(\delta_{\mu\nu}) =\text{diag}(+1,+1,+1,+1)$. Our results can be easily adapted to Lorentzian signature and an expansion of the metric around around the flat Minkowski metric $(\eta_{\mu\nu}) =\text{diag}(-1,+1,+1,+1)$. For notational convenience, we will work in the Euclidean version of the Fourier space. 
},
\be\label{eq:ExpansionMetric}
g_{\mu\nu} = \delta_{\mu\nu} + h_{\mu\nu}\,,
\ee
where $h_{\mu\nu}$ denotes the symmetric rank-two tensor of metric perturbations. The latter can be combined with two powers of momenta to form four independent contractions\footnote{For notational convenience we adopt the following notation for the Fourier-transformed Lagrangians: A term $ \phi_{\mu\cdots} K(p^2)^{\mu\cdots\nu\cdots}  \psi_{\nu\cdots}$, with $\phi_{\mu\cdots},\psi_{\mu\cdots}$ field variables and $K(p^2)^{\mu\cdots\nu\cdots} $ a quadratic polynomial in $p^\mu$, stands for $\tfrac{1}{2}\phi_{\mu\cdots}(p) K(p^2)^{\mu\cdots\nu\cdots}  \psi_{\nu\cdots}(-p)  + \tfrac{1}{2}\psi_{\mu\cdots}(p) K(p^2)^{\mu\cdots\nu\cdots}  \phi_{\nu\cdots}(-p) $.\label{Notation} }
 $ h_{\mu\nu}h^{\mu\nu}p^2$, $h_{\mu\rho} h_{\nu}^{\rho}p^\mu p^\nu $, $h h_{\mu \nu}p^\mu p^\nu $, $h^2 p^2 $, where $h=h_{\mu\nu}\delta^{\mu\nu}$ denotes the trace of $h_{\mu\nu}$. All four of these terms can occur with free coupling constants in the most general second-order quadratic Lagrangian for the field $h_{\mu\nu}$. However, demanding invariance of the action under linearized diffeomorphisms parametrized by the vector field $\xi^\mu$,
\be\label{DiffeomorphismTransformationMetric}
h_{\mu\nu} \to h_{\mu\nu} + p_\mu \xi_\nu + p_\nu \xi_\mu\,,
\ee
uniquely fixes three of these couplings as functions of the remaining one. The most general diffeomorphism-invariant second-order quadratic Lagrangian for metric gravity takes the form
\be\label{eq:LagrangianEHDiffInvariant}
\mathcal{L}_{\text{diff.inv.}}(h_{\mu\nu})\propto  \frac{1}{2} h_{\mu\nu}h^{\mu\nu}p^2 -\frac{1}{2}h^2 p^2 - h_{\mu\rho} h_{\nu}^{\rho}p^\mu p^\nu  + h h_{\mu \nu}p^\mu p^\nu = \mathcal{L}_{\text{EH}}(h_{\mu\nu}) \,.
\ee
Thus, the condition of diffeomorphism invariance leads to the linearized Einstein-Hilbert action up to a global rescaling.

One of the goals of the present paper is to apply the same procedure to actions based on area metrics.  This allows us to identify a space of covariant kinetic terms for area metrics in Sec.~\ref{Sec:AreaMetrics} and then the space of diffeomorphism-invariant linearized area-metric actions in Sec.~\ref{Sec:DiffeomorphismInvarianceCondition}. We show that a certain subset of these actions has a degenerate kinetic term --- this subset coincides with the actions constructed from modified Plebanski theory in \cite{BorDitt}. The degeneracy of the kinetic term can be further understood and explored from two different viewpoints. First, in Sec.~\ref{Sec:EffectiveLengthMetricActions}, integrating out the non-length degrees of freedom, we construct effective length metric actions. We will see that the theories with degenerate kinetic terms leads to ghost-free propagators for the effective length-metric action.  This subset is described by two coupling parameters --- one parameter parametrizes parity-violating terms, the other parametrizes the mass for the degrees of freedom which are not induced by a length metric. 

A different viewpoint that we pursue is to perform the canonical analysis of the area metric theories. This gives an alternative understanding of the fact that a subclass of actions leads to a ghost-free propagator for the effective theory. We perform the canonical analysis switching to the Lorentzian version of the theory in Sec.~\ref{Sec:WickRotation}.  We find, that the Lorentzian action --- but not the Euclidean action --- contains propagating degrees of freedom with negative definite kinetic and mass terms. This is a possible source of instability already in the classical theory. Nevertheless, a careful analysis in Sec.~\ref{Sec:HamiltonianAnalysis} of the equations of motion, done using the mode expansion, shows that the linearized dynamics is stable, despite having positive and negative definite kinetic terms in the action, and a coupling between these terms. This means that the problem of viability of the Lorentzian area metric gravity actions becomes a non-linear problem. We further comment on all these issues in the discussion section \ref{Sec:Discussion}.

  \section{Area metrics and covariant area-metric actions}\label{Sec:AreaMetrics}
  
  \subsection{Definition and symmetries}
  
An area metric $G$ at a point $p$ on a smooth manifold $M$ is a rank-four covariant tensor $G: (T_p M)^{\otimes 4} \to \mathbb{R}$ with the following symmetries,
\be\label{eq:AreaMetricSymmetries}
G_{\mu\nu\rho\sigma} = - G_{\nu\mu\rho\sigma} = G_{\rho\sigma\mu\nu}\,.
\ee
Therefore an area metric can be regarded as a metric for bi-vectors and defines the linear map
\be
G:  \Lambda^2 TM \to (\Lambda^2 TM)^*\,, \quad B^{\mu\nu} \mapsto G_{\mu\nu\rho\sigma}B^{\rho\sigma}\,,
\ee
where existence of the inverse $G^{-1}$ guarantees an unambiguous raising and lowering of bi-vector indices. A given area metric can be decomposed uniquely into the sum of two parts: the first part is called cyclic and satisfies the algebraic Bianchi identity
\be \label{eq:CyclicityCondition}
G_{\mu[\nu\rho\sigma]} = 0 \quad \Leftrightarrow \quad G_{\mu\nu\rho\sigma} \epsilon^{\mu\nu\rho\sigma} = 0\, \q ,
\ee 
where $\epsilon^{\mu\nu\rho\sigma}$ denotes the Levi-Civita density and the equivalence holds provided the algebraic symmetries~\eqref{eq:AreaMetricSymmetries} are satisfied.  The second part is a four-form and is thus totally anti-symmetric. 

Cyclic area metrics are area metrics for which this four-form part vanishes. They have therefore the same algebraic symmetries as the Riemann tensor. In the following we will restrict to cyclic area metrics.

This restriction can be motivated as follows~\cite{BorDitt}. The components of a length metric can be reconstructed by only measuring the lengths of basis vectors and of sums of basis vectors in the tangent space, without measuring angles. Similiarly, the components of an area metric can be reconstructed from measuring areas of parallelograms, however, only up to combinations involving cyclic sums of area metric components of the form $G_{\mu[\nu\rho\sigma]}$. The cyclicity condition~\eqref{eq:CyclicityCondition} sets these terms to zero, such that a cyclic area metric can be recovered by measuring only areas without an independent measurement of dihedral angles being necessary.\\

Every length metric $g_{\mu\nu}$ induces a cyclic area metric by the definition
\be
G^{\text{(ind.)}}_{\mu\nu\rho\sigma}(g) = g_{\mu\rho}g_{\nu\sigma} - g_{\mu\sigma}g_{\nu\rho}\,.
\ee

However, a general cyclic area metric in four dimensions has twenty independent components and therefore twice as many as a length metric in four dimensions. Thus, not every area metric is induced by a length metric and the notion of area-metric spacetimes provides a much more general concept than length metric spacetimes.\\

\subsection{Parametrization of area-metric perturbations and their irreducible components}\label{Sec:Param}

Analogous to the expansion of the length metric around a flat Euclidean background~\eqref{eq:ExpansionMetric}, in the following we consider an area metric expanded around a configuration induced by the flat Euclidean metric,
\be \label{eq:ExpansionAreaMetric}
G_{\mu\nu\rho\sigma}= G^{\text{(ind.)}}_{\mu\nu\rho\sigma}(\delta) +a_{\mu\nu\rho\sigma}= 2\delta_{\mu[\rho}\delta_{\sigma] \nu} +a_{\mu\nu\rho\sigma}\,,
\ee
where $a_{\mu\nu\rho\sigma}$ denotes the (cyclic) area metric perturbations. Our goal in the next sections will be to construct the full set of possible kinetic terms of second order in the momenta and combine these into an action whose free parameters will later be constrained through the requirement of diffeomorphism invariance. Additionally, we will consider mass terms for a subset of the degrees of freedom of the area metric.\\

The twenty components of the area metric perturbation $a_{\mu\nu\rho\sigma}$ can be decomposed into irreducible representations of ${\rm SO}(4)$, which stabilizes the background metric $\delta_{\mu\nu}$. This decomposition is the same as the familiar decomposition of the Riemann curvature tensor.
 In four dimensions, the latter decomposes into the Ricci scalar, the trace-free part of Ricci tensor, and the self- and anti-self dual parts of the Weyl curvature. We write, similarly
\be\label{a-decomp}
a_{\mu\nu\rho\sigma} = h \delta_{\rho[\mu} \delta_{\nu]\sigma} +2( \tilde{h}_{\rho[\mu} \delta_{\nu]\sigma} - \tilde{h}_{\sigma[\mu} \delta_{\nu]\rho}) + w^+_{\mu\nu\rho\sigma} + w^-_{\mu\nu\rho\sigma} \q .
\ee
Here $\tilde{h}_{\mu\nu}$ is tracefree, $\tilde{h}_{\mu\nu}\delta^{\mu\nu}=0$, and $w^\pm_{\mu\nu\rho\sigma}$ are both tracefree, $w^\pm_{\mu\nu\rho\sigma} \delta^{\mu\rho}=0$, and satisfy the self-duality equations (\ref{ESD}).
\be\label{ESD}
\frac{1}{2} \epsilon_{\mu\nu}{}^{\alpha\beta} w^\pm_{\alpha\beta\rho\sigma}= \pm w^\pm_{\mu\nu\rho\sigma}.
\ee

The general irreducible representation of ${\rm SO}(4)$ is of the type $(j,j')$, where $j,j'$ are both integers or both half-integers. It is of dimension ${\rm dim}(j,j') = (2j+1)\times(2j'+1)$. The representations that appear in (\ref{a-decomp}) are 
\be\label{irreps-a}
a \in (0,0) \oplus (1,1) \oplus (2,0) \oplus (0,2).
\ee

We will now proceed to determine the most general diffeomorphism invariant action to second order in the area metric perturbations, and to second order in derivatives. We assume that the area metric induced by the flat length metric is a solution of the non-perturbative area metric. We can thus set the linear terms (modulo boundary terms) to zero and only need to consider terms quadratic in fluctuations. In the next section we determine all possible  kinetic terms quadratic in derivatives. Terms in which the two derivatives contract with each other to the Laplacian can be generalized to include mass parameters. 

\subsection{Kinetic terms for the area-metric perturbation}\label{Sec:KinTerms}

We now proceed to determine all possible kinetic terms, quadratic in $a$ and of second order in derivatives, that can be written. It is standard to perform such an analysis in the momentum space, and so we will have two factors of the momentum $p^{\mu} p^\nu$ in each kinetic term. 

It is straightforward to write down all possible terms where the factors of the momenta contract between themselves to produce $p^2$. Indeed, these terms are just
\be
p^2 h^2, \qquad p^2 (\tilde{h}_{\mu\nu})^2, \qquad p^2 (w^+_{\mu\nu\rho\sigma})^2, \qquad p^2 (w^-_{\mu\nu\rho\sigma})^2,
\ee
which are built from the squares of each of the irreducible component. 

To understand the possible terms that do not involve $p^2$, we note that each factor of the momentum is in the vector $(1/2,1/2)$ representation of ${\rm SO}(4)$. The product $p^\mu p^\nu$ is in the symmetric part of the tensor product
\be
p\otimes p\in (1/2,1/2)\otimes_S (1/2,1/2) = (1,1)\oplus(0,0) \ni (p\otimes p)_{tf} \oplus p^2.
\ee
The second factor is where $p^2$ resides, and the first factor is the tracefree part of $p^\mu p^\nu$, which we denote by $(p\otimes p)_{tf}$. Given that we already described all possible terms involving $p^2$, we need to understand all possible singlets that can be constructed from the representation $(1,1)$, coming from the tracefree part of $p^\mu p^\nu$, and two copies of the representations appearing in (\ref{irreps-a}). 

Firstly, because $(0,0)\otimes (1,1)=(1,1)$, it is clear that the singlet representation $h$ can only appear in the combination
\be\label{h-tilde-h}
h p^\mu p^\nu \tilde{h}_{\mu\nu},
\ee
which is one of the two terms not involving $p^2$ that appear in the linearisation of the Einstein-Hilbert action. 

Secondly, let us determine what $\tilde{h}_{\mu\nu}$, if combined with $(p \otimes p)_{tf}$, can couple to. We have the following decomposition
\be
\tilde{h}\otimes (p \otimes p)_{tf} \in (1,1)\otimes(1,1) = (2,2)\oplus (1,1) \oplus (0,0)\oplus  (2,1)\oplus (1,2) \oplus (2,0) \oplus (0,2) \oplus (1,0)\oplus (0,1) ,
\ee
where every representation appears with multiplicity one. To find possible kinetic terms we have to tensor these representations with the representations appearing in (\ref{irreps-a}). The kinetic term is a singlet, the singlet appears with multiplicity one in the tensor product $(j,j') \otimes (j,j') =(0,0)\oplus \cdots$. 
It is clear that all the four representations from the list (\ref{irreps-a}) appear here, and so there are four corresponding kinetic terms. The kinetic term of the schematic type $\tilde{h} pp h$ already appeared in (\ref{h-tilde-h}). The kinetic term of the type $\tilde{h} pp \tilde{h}$ is given by
\be
(p^\mu \tilde{h}_{\mu\nu})^2,
\ee
and is the second of the terms not involving $p^2$ that appears in the linearisation of the Einstein-Hilbert action. The terms involving $w^\pm$ are
\be\label{w-tilde-h}
\tilde{h}^{\mu\rho} p^\nu p^\sigma w^\pm_{\mu\nu\rho\sigma}.
\ee

It remains to see that there are no new invariant terms of the schematic type $(pp)_{tf} w^+ w^+$, $(pp)_{tf}  w^- w^-$ or $(pp)_{tf}  w^+ w^-$ that can be constructed. Let us consider
\be
w^+ \otimes (p\otimes p)_{tf} \in (2,0)\otimes (1,1) = (3,1)\oplus (2,1)\oplus (1,1). 
\ee
The only representation from the list (\ref{irreps-a}) that appears here is $(1,1)$. But the resulting kinetic term has already been listed in (\ref{w-tilde-h}). Thus no kinetic terms of the type $(pp)_{tf}  w^+ w^+$, $(pp)_{tf}  w^- w^-$ or $(pp)_{tf}  w^+ w^-$ are possible.

To summarise, we see that there are just 8 possible kinetic terms that can be constructed for the area metric. They are best described by decomposing the area metric into its ${\rm SO}(4)$ irreducible parts. Then four of these kinetic terms are those already present in the linearisation of the Einstein-Hilbert action:
\be
p^2h^2, \qquad p^2 (\tilde{h}_{\mu\nu})^2, \qquad hp^\mu p^\nu \tilde{h}_{\mu\nu}, \qquad (p^\mu \tilde{h}_{\mu\nu})^2.
 \ee
 The four new terms involving $w^\pm$ are
\be
p^2 (w^+_{\mu\nu\rho\sigma})^2, \qquad p^2 (w^-_{\mu\nu\rho\sigma})^2, \qquad \tilde{h}^{\mu\rho} p^\nu p^\sigma w^+_{\mu\nu\rho\sigma}, \qquad \tilde{h}^{\mu\rho} p^\nu p^\sigma w^-_{\mu\nu\rho\sigma}.
\ee

\subsection{The general area-metric Lagrangian}

In the last section we have split the area-metric perturbation $a_{\mu\nu\rho\sigma}$ into four parts, $h$, $\tilde{h}_{\mu\nu}$ and $w^{\pm}_{\mu\nu\rho\sigma}$. The first two parts combine into the length-metric perturbation
\ba
h_{\mu\nu}=\tilde h_{\mu\nu}+\tfrac{1}{4} \delta_{\mu\nu}h \quad ,
\ea
and encode 10 of the 20 degrees of freedom of the area metric perturbation. 

The Weyl-curvature like  parts $w^{\pm}_{\mu\nu\rho\sigma}$ include each 5 degrees of freedom.  These $(2\times 5 )$ degrees of freedom can be encoded into a pair of trace-free  and symmetric matrices $\chi_{ij}^\pm$ of space-time scalars, where $i,j=1,2,3$ are internal indices which can be raised and lowered with $\delta^{ij}$ and $\delta_{ij}$, respectively.  
The $\chi^\pm_{ij}$ and  $w^\pm_{\mu\nu\rho\sigma}$ are related by
\ba
(\chi^\pm)^{ij}\,=\,    \tfrac{1}{2}    \mathbb{P^\pm}^{ij}_{\mu\nu\rho\sigma} a^{\mu\nu\rho\sigma}\,=\,     \tfrac{1}{2}    \mathbb{P^\pm}^{ij}_{\mu\nu\rho\sigma} (w^\pm)^{\mu\nu\rho\sigma} ,\q \q\q\q w^\pm_{\mu\nu\rho\sigma}=2\mathbb{P^+}^{ij}_{\mu\nu\rho\sigma} \chi^+_{ij},
\ea
where
\ba
\mathbb{P^\pm}^{ij}_{\mu\nu\rho\sigma} & =&  \tfrac{1}{8} \left( {\Sigma^\pm}^i_{\mu\nu}    {\Sigma^\pm}^j_{\rho\sigma}   + {\Sigma^\pm}^j_{\mu\nu}    {\Sigma^\pm}^i_{\rho\sigma}     \right) -\tfrac{1}{12} \delta^{ij}  {\Sigma^\pm}^{i'}_{\mu\nu}    {\Sigma^\pm}^{j'}_{\rho\sigma}\delta_{i'j'}  \,, \quad  \text{with} \nn\\
{\Sigma^\pm}^i_{\mu\nu} &=&  \pm ({\delta}^0_\mu {\delta}^i_\nu-{\delta}^0_\nu {\delta}^i_\mu)+{\epsilon^i}_{jk}{\delta}^j_\mu {\delta}^k_\nu .
\ea
The ${\Sigma^\pm}^i_{\mu\nu}$ coincide with the Plebanski (selfdual  or anti-selfdual) two-form evaluated on a flat (Euclidean) background. They thus satisfy the (anti-) selfduality condition (\ref{ESD}).

The couplings involving the $w^\pm_{\mu\nu\rho\sigma}$ then translate\footnote{Here we use the identity $\mathbb{P^\pm}^{ij} \circ \mathbb{P^\pm}^{i'j'}= \delta^{i(i'} \delta^{j')j} -\tfrac{1}{3} \delta^{ij} \delta^{i'j'} $.} as follows:
\ba
(\omega^\pm_{\mu\nu\rho\sigma})^2&=& 4 (\chi^\pm_{ij})^2     
\ea
and
\ba\label{hchiCoupling}
\tilde{h}^{\mu\rho}p^\nu p^\sigma w^\pm_{\mu\nu\rho\sigma} &=&2 h^{\mu\rho} p^\nu p^\sigma \mathbb{P^\pm}^{ij}_{\mu\nu\rho\sigma} \chi^\pm_{ij}\,=\, \tfrac{1}{2}h^{\mu\rho} p^\nu p^\sigma {\Sigma^\pm}^i_{\mu\nu}    {\Sigma^\pm}^j_{\rho\sigma}  \chi^\pm_{ij} \q .
\ea
 Here we used that  $\chi^\pm_{ij}$ is symmetric and trace-free. One can then also find that  ${\Sigma^\pm}^i_{\mu\nu}{\Sigma^\pm}^j_{\rho\sigma}  \chi^\pm_{ij}\delta^{\mu\rho}=0$, which allows us to replace the trace-free $\tilde h^{\mu\rho}$ with $h^{\mu\rho}$. 
 
 Indeed, it is only the trace-free transverse parts of the length metric perturbation $h_{\mu\nu}$, which can couple to the $w^+_{\mu\nu\rho\sigma}$ and $w^-_{\mu\nu\rho\sigma}$ part of the area-metric perturbations respectively. We can isometrically embed the $\chi_{ij}^\pm$ into the space of symmetric, transverse and traceless rank-two space-time tensors as follows:
\ba\label{eq:ChiSpacetimeIndices}
 \chi^\pm_{\mu \rho} \equiv {\Sigma^\pm}^i_{\mu\nu}    {\Sigma^\pm}^j_{\rho\sigma}   \frac{p^\nu p^\sigma} {p^2}  \chi^\pm_{ij} \equiv {E^{\pm}}^{ ij}_{\mu\rho} \chi^{\pm}_{ij}\, .
 \ea
Thus the following equations hold: $\chi^\pm_{\mu\nu}p^\mu = 0$, $\chi^{\pm }_{\mu\nu}\delta^{\mu\nu} = 0$ and ${\chi^\pm}_{\mu\nu}\chi^{\pm \mu\nu}={\chi^\pm}_{ij}\chi^{\pm ij} $.

~\\

In Section \ref{Sec:KinTerms} we constructed 8 possible kinetic terms for the area metric. We will also allow in our general Lagrangian mass terms for the $\chi^\pm_{ij}$ degrees of freedom.  The most general quadratic Lagrangian, with second order derivatives, is therefore given by 
 \ba\label{eq:LagrangianHChi}
 \mathcal{L}(h_{\mu\nu},\chi^+_{\mu\nu},\chi^-_{\mu\nu}) &=& A_0 h_{\mu\nu}h^{\mu\nu}p^2 + A_1 h^2 p^2 +A_2 h_{\mu\rho} h_{\nu}^{\rho}p^\mu p^\nu  +A_3 h h_{\mu \nu}p^\mu p^\nu   \nn\\
 &+&\frac{\alpha_+}{2} h_{\mu\nu}\chi\indices{^+^{\mu\nu}}p^2 +  \frac{\alpha_-}{2} h_{\mu\nu}\chi\indices{^-^{\mu\nu}}p^2 +  \frac{\beta_+}{4}\chi^+_{\mu\nu} \chi\indices{^+^{\mu\nu}}p^2 + \frac{\beta_-}{4}\chi^-_{\mu\nu} \chi\indices{^-^{\mu\nu}}p^2 +\frac{m_+^2}{4} \chi^+_{\mu\nu} \chi\indices{^+^{\mu\nu}} + \frac{m_-^2}{4} \chi^-_{\mu\nu} \chi\indices{^-^{\mu\nu}} \,.\nn\\
 \ea

This Lagrangian (\ref{eq:LagrangianHChi}) can be translated into a quadratic form in area-metric perturbations via the following relations (which make use of the conventions of Footnote \ref{Notation})
 \ba\label{eq:BasicRelations}
 a_{\alpha\beta\gamma\delta}a^{\alpha\beta\gamma\delta}p^2 &=& 8 h_{\mu\nu}h^{\mu\nu}p^2 + 4 h^2p^2 + 4 \chi^{+}_{ij}\chi^{+ ij} p^2+  4 \chi^{-}_{ij}\chi^{- ij}p^2\,,\nn\\
 a\indices{^{\alpha\beta}_\alpha ^\gamma}a\indices{_\beta^\delta_\gamma_\delta}p^2 &=& 4  h_{\mu\nu}h^{\mu\nu} p^2+ 8 h^2 p^2\,,\nn\\
  a\indices{^\alpha^\beta _\alpha_\beta} a\indices{^\gamma^\delta _\gamma_\delta} p^2&=& 36 h^2 p^2\,,\nn\\
  a\indices{_\alpha_\beta^\mu^\nu}a^{\alpha\beta\gamma\delta}\epsilon_{\gamma\delta\mu\nu}p^2 &=&  8 \chi^{+}_{ij}\chi^{+ ij} p^2-  8 \chi^{-}_{ij}\chi^{- ij}p^2\,,\nn\\
  a\indices{_\alpha^\gamma_\beta^\delta}a\indices{_\gamma^\mu_\delta_\mu}p^\alpha p^\beta &=& 2 h_{\mu\nu}h^{\mu\nu}p^2 + h^2p^2 -4 h_{\mu\rho}h^\rho_\nu p^\mu p^\nu +4 h h_{\mu\nu}p^\mu p^\nu +   p^2{ E^{+}}^{ij}_{\mu\nu} \chi^+_{ij} h^{\mu\nu}  
    +  p^2{ E^{-}}^{ij}_{\mu\nu} \chi^-_{ij} h^{\mu\nu} \, ,  
   \nn\\
  a\indices{_\alpha^\gamma_\beta_\gamma} a\indices{^\delta^\mu_\delta_\mu}p^\alpha p^\beta &=&  6 h^2p^2 +12 h h_{\mu\nu}p^\mu p^\nu           \, ,\nn\\
  a\indices{_\alpha^\gamma_\gamma^\delta}a\indices{^\mu_\beta_\delta_\mu}p^\alpha p^\beta &=&  h^2p^2 +4 h_{\mu\rho}h^\rho_\nu p^\mu p^\nu +4 h h_{\mu\nu}p^\mu p^\nu\,,\nn\\
  a\indices{_\alpha^\gamma^\delta^\lambda}a\indices{_\gamma ^\mu _\mu ^\nu}\epsilon_{\beta\delta\nu\lambda}p^\alpha p^\beta &=&  2p^2 { E^{+}}^{ij}_{\mu\nu} \chi^+_{ij} h^{\mu\nu}  
    - 2p^2 { E^{-}}^{ij}_{\mu\nu} \chi^-_{ij} h^{\mu\nu}\, . 
 \ea
Note that in terms of the area-metric perturbations we have two contractions which include the Levi-Civita tensor density, and are hence parity-breaking.  (Any terms with more than one Levi-Civita tensor density can be rewritten into terms containing one or zero such densities.) These indeed lead to differences between terms involving the $\chi^+$ and the $\chi^-$ fields.

~\\~\\
 {\bf Remark:}  In the covariant version of constructive gravity the work \cite{ConGrav2} identifies 37 possible terms quadratic in area metric fluctuations. 
 For these 37 terms one does however allow for the acyclic part $\lambda \epsilon_{\mu\nu\rho\sigma}$, with $\lambda$ a scalar density, for the area metrics. 
 In \cite{ConGrav2} one also distinguishes terms which are related by integrations by part. The counting of 37 does split into 6 terms without derivatives and 31 quadratic in derivatives. Allowing for acyclic area metrics, we obtain additional couplings $(p^2)\lambda^2, (p^2)\lambda h$ and $\tilde h_{\mu\nu}p^\mu p^\nu\lambda$. Identifying terms related by integrations by part we can then match all 31 terms of \cite{ConGrav2} with our 8 contractions given in (\ref{eq:BasicRelations}) and  the 3 terms involving $\lambda$. We do match the six terms without derivatives one-to-one, these are given by $h^2, (\tilde h_{\mu\nu})^2$ and $(\chi_+^{ij})^2, (\chi_-^{ij})^2$ as well as $\lambda^2,\lambda h$. Thus, our results here are not in conflict with those in \cite{ConGrav2}, but are obtained in a much more straighforward way. 
 
~\\~\\
 {\bf Remark:} 
Let us remark that the parametrization of the area-metric fluctuations in terms of a length-metric perturbations $h_{\mu\nu}$ and the 10 $\chi^i_\pm$ fields can be extended to the non-linear theory \cite{BorDitt}. To that end, one starts from (non-chiral) Plebanski theory, which features $\text{so}(4)$-valued two-forms $B^{IJ}_{\mu\nu}$, with $I,J=0,\ldots,3$ as configuration variables. Via the decomposition of $\text{so}(4)$ into selfdual and anti-selfdual parts $\text{so}(3)\oplus \text{so}(3)$, we obtain two $\text{so}(3)$ valued two-forms $(B_\pm)^i_{\mu\nu}$. These two-forms can be parametrized by two uni-modular $3\times 3$ matrices $(b_\pm)^i_j$ and two tetrads $(e_\pm)^I_\mu$, see \cite{FreidelMod}.  Forcing the two tetrads to be equal one reduces the number of $SO(4)$ invariant degrees of freedom encoded in the $B$-fields from 30 to 20. This allows for defining a cyclic area metric \cite{BorDitt}, which is parametrized by the length metric (obtained from the tetrad field), and two fields of uni-modular $3\times 3$ matrices. Expanding these data around a flat background, one finds that the perturbation can be parametrized by length-metric fluctuations $h_{\mu\nu}$ and 10 scalar fields $\chi^{ij}_\pm$ \cite{BorDitt}.

  \section{Requirement of diffeomorphism invariance}\label{Sec:DiffeomorphismInvarianceCondition}

In this section we will illustrate how the requirement of invariance of the Lagrangian~\eqref{eq:LagrangianHChi} under linearized diffeomorphism transformations further restricts the number of free parameters.

Linearized diffeomorphisms parametrized by the vector field $\xi^\mu$ act on the symmetric rank-two tensor of metric perturbations $h_{\mu\nu}$ and on the space-time scalars $\chi^\pm_{ij}$ as follows
\be\label{eq:DiffeomorphismTransformation}
h_{\mu\nu} \to h_{\mu\nu} + p_\mu \xi_\nu + p_\nu \xi_\mu\,,\quad \chi^\pm_{ij}\to \chi^\pm_{ij}\,.
\ee
Demanding invariance of the Lagrangian~\eqref{eq:LagrangianHChi} under linearized diffeomorphisms~\eqref{eq:DiffeomorphismTransformation} fixes the part quadratic in the field $h_{\mu\nu}$ to a multiple of the Einstein-Hilbert action,
\footnote{Here the spin-2 projector ${}^2\!P$ which projects onto the symmetric, transverse and traceless tensor modes and the spin-0 projector ${}^0\!P$ which projects onto the symmetric and transverse trace tensor modes are defined in four spacetime dimensions as
	\be\label{eq:LagrangianH}
	{}^0 \!P_{\mu\nu\rho\sigma}  =  \frac{1}{3} P^{\perp}_{\mu\nu} P^{\perp}_{\rho\sigma}\,,\quad \quad 
	{}^2\! P_{\mu\nu\rho\sigma}= \frac{1}{2}(P^{\perp}_{\mu\rho}P^{\perp}_{\rho\sigma}+P^{\perp}_{\mu\sigma}P^{\perp}_{\nu\rho}) -  \frac{1}{3} P^{\perp}_{\mu\nu} P^{\perp}_{\rho\sigma} \,,\quad\text{where} \quad
	P^{\perp}_{\mu\nu} =\delta_{\mu\nu}-\frac{p_\mu p_\nu}{p^2}\,.
	\ee
}
\be
\mathcal{L}_{\text{diff.inv.}} (h_{\mu\nu})  = A\, \mathcal{L}_{\text{EH}}(h_{\mu\nu}) =\frac{A}{2}\,h_{\mu\nu}\big({}^{2}\!P^{\mu\nu\rho\sigma} - 2\,{}^0\! P^{\mu\nu\rho\sigma}\big)p^2h_{\rho\sigma}\,.
\ee
On the other hand, the part of the Lagrangian which is quadratic in the fields $\chi^{\pm}_{ij}$, is invariant as these do not transform under diffeomorphisms. Moreover, these fields only couple to the transverse and traceless part of the length metric perturbations, such that the coupling terms between $\chi^\pm_{ij}$ and $h_{\mu\nu}$ are left invariant under the diffeomorphism transformation~\eqref{eq:DiffeomorphismTransformation}. As a result, demanding diffeomorphism invariance leaves us with the four free parameters $\alpha_\pm$ and $\beta_\pm$ in the notation of~\eqref{eq:LagrangianHChi}, plus one additional global rescaling parameter denoted by $A$ in front of the Einstein-Hilbert term~\eqref{eq:LagrangianH}. Allowing mass terms for the $\chi^\pm$-fields the most general second-order quadratic diffeomorphism-invariant Lagrangian for $h_{\mu\nu}\,, \chi^{\pm}_{\mu\nu}$ (and thus for the area metric $a_{\mu\nu\rho\sigma}$) can be written in the form
\be\label{eq:LagrangianHChiDiffInvariant}
\mathcal{L}_{\text{diff.inv.}}(h_{\mu\nu},\chi^+_{ij},\chi^-_{ij}) = A  \mathcal{L}_{\text{EH}}(h_{\mu\nu}) + \frac{1}{4}\sum_{\pm} \left(2\alpha_\pm h_{\mu\nu} {E^{\pm}}^{\mu\nu}_{ij} \chi^{\pm ij} p^2+ \beta_\pm\chi^{\pm ij}\chi^\pm_{ij}p^2 + m_\pm^2 \chi^{\pm ij}\chi^\pm_{ij} \right)\,.
\ee

We close this section with a number of remarks:
\begin{itemize}
\item If $m^2_\pm>0$, the fluctuations of $\chi^\pm$ will be suppressed by these mass terms. We can integrate out the  $\chi^\pm$ fields and obtain to leading order in a derivative expansion $A$ times the Einstein-Hilbert term (see below). Thus, if we wish to recover (linearized) general relativity, we have to choose $A=1$. If one is only interested in the theory after integrating out the $\chi^\pm$ fields, one can redefine the $\chi^\pm$ fields by a rescaling (and the sign of $\alpha$). This absorbs two couplings. We are left with the following combination of four coupling constants:
$M_\pm^2 = m_\pm^2/\beta_\pm$ and $\rho_\pm = \alpha_\pm^2/\beta_\pm$. 
\item For a parity-preserving Lagrangian we need to set $\alpha_+=\alpha_-$ and $\beta_+=\beta_-$ and $m^2_+=m^2_-$.
\item The Lagrangian (\ref{eq:LagrangianHChiDiffInvariant}) features an additional shift symmetry for special values of the coupling constants:  Assume the masses $m_\pm^2$ are vanishing  and the couplings  satisfy
\ba\label{shiftcond}
\frac{\alpha_+^2}{\beta_+}+\frac{\alpha_-^2}{\beta_-}\,=\, 2A \quad ,
\ea
with $A>0$ and $\beta_\pm>0$.
The Lagrangian (\ref{eq:LagrangianHChiDiffInvariant}) (without mass terms) can then be written as a sum of two squares
\ba
\mathcal{L}_{\text{kin}}\,=\,\frac{p^2}{4}\sum_\pm  \left( \frac{\alpha_\pm}{\sqrt{\beta_\pm}} h_{\mu\nu} + \,\sqrt{\beta_\pm} E_{\mu\nu}^{\pm\, ij} \chi^\pm_{ij}\right)\big({}^{2}\!P^{\mu\nu\rho\sigma} - 2\,{}^0\! P^{\mu\nu\rho\sigma}\big)  \left( \frac{\alpha_\pm}{\sqrt{\beta_\pm}} h_{\rho\sigma} + \,\sqrt{\beta_\pm} E_{\rho\sigma}^{\pm\, ij} \chi^\pm_{ij}\right)   \quad .
\ea
This leads to a 5-parameter\footnote{
This shift symmetry (\ref{shiftsym} can be expressed in more symmetric form as
\ba
\chi^+_{ij} &\rightarrow &\chi^+_{ij} + \frac{\alpha_+}{\beta_+}\left( \zeta^+_{ij}  +
\delta_{im}\delta_{jn}E^{+mn}_{\mu\nu}\delta^{\mu\rho}\delta^{\nu\sigma} E^{-kl}_{\rho\sigma} \zeta^-_{kl}
\right)\,,\quad \nn\\
\chi^-_{ij}& \rightarrow& \chi^-_{ij} + 
 \frac{\alpha_-}{\beta_-}\left( \zeta^-_{ij}  +
\delta_{im}\delta_{jn}E^{-mn}_{\mu\nu}\delta^{\mu\rho}\delta^{\nu\sigma} E^{+kl}_{\rho\sigma} \zeta^+_{kl}
\right)
\,,\quad \nn\\
\quad h_{\mu\nu} &\rightarrow&  h_{\mu\nu}- E^{+\,ij}_{\mu\nu} \zeta^+_{ij} - E^{-\,ij}_{\mu\nu} \zeta^-_{ij} \, .
\ea
Here we have five parameters $\zeta^+_{ij}$ and five parameters $\zeta^-_{ij}$ appearing, but one can show that there are five redundancies between these parameters.
}
 gauge symmetry, which is in addition to the linearized diffeomorphisms: 
\ba\label{shiftsym}
\chi^+_{ij} \rightarrow \chi^+_{ij} + \zeta_{ij} \,,\quad
\chi^-_{ij} \rightarrow \chi^-_{ij} + \frac{\alpha_-\beta_+}{\alpha_+\beta_-} \delta_{im}\delta_{jn} E^{-mn}_{\mu\nu}\delta^{\mu\rho}\delta^{\nu\sigma} E^{+kl}_{\rho\sigma} \zeta_{kl} \,,\quad
\quad h_{\mu\nu} \rightarrow  h_{\mu\nu}-\frac{\beta_+}{\alpha_+}E^{+\,ij}_{\mu\nu} \zeta_{ij} \, ,
\ea
where $\zeta_{ij}$ is symmetric and trace-less. To see this one uses
\ba
E^{-ij}_{\mu\nu} \delta_{ik}\delta_{jl} E^{-kl}_{\rho\sigma}  \delta^{\rho\kappa} \delta^{\sigma\kappa} E^{+mn}_{\kappa\lambda} \,=\, E^{+mn}_{\mu\nu} \, ,
\ea
as $E^{-ij}_{\mu\rho} \delta_{ik}\delta_{jl} E^{-kl}_{\rho\sigma}$ equates to the ${}^2\! P$ projector, which acts as an identity on $E^{+mn}$.  Adding masses $m^2_\pm$  breaks this shift symmetry (\ref{shiftsym}). In this case one has still a degenerate kinetic term. That is, the corresponding quadratic form has dimension $20\times 20$ but features 9 null vectors --- 4 result from the linearized diffeomorphism invariance and 5 from the shift symmetry (\ref{shiftsym}).

This degeneracy of the kinetic term appears for  area-metric actions derived from  the  modified  Plebanski action, see \cite{FreidelMod,Krasnov2,Spez,BorDitt}. It can be seen as a remnant of a shift symmetry for the two-form $B$-field appearing in the Plebanski action. The Plebanski action is a sum of the $BF$ action, which is invariant under such a shift symmetry, and constraint terms, which break part of these symmetries.
\end{itemize}

 \section{Effective length-metric actions and propagators}\label{Sec:EffectiveLengthMetricActions}

   Here we will integrate out the $\chi^\pm$ fields from the Lagrangian~\eqref{eq:LagrangianHChiDiffInvariant} and in this way define an effective Lagrangian of the length metric perturbations only, which captures the effects resulting from the additional degrees of freedom in the area metric. 
 
Varying the Lagrangian~\eqref{eq:LagrangianHChiDiffInvariant} with respect to the fields $\chi^{\pm}_{ij}$ using the definition~\eqref{eq:ChiSpacetimeIndices}, we find
\be
\pdv{\mathcal{L}_{\text{diff.inv.}}}{\chi^\pm_{ij}} = 0 \quad \Rightarrow \quad \chi^\pm_{ij} = -\frac{\rho_\pm}{\alpha_\pm}\qty(\frac{1}{p^2 +M^2_\pm}) p^2 h_{\mu\nu}E^{\pm \mu\nu}_{ij}\,,
\ee
where $\rho_\pm=\alpha^2_\pm/\beta_\pm$ and $M^2_\pm=m^2_\pm/\beta_\pm$.
Reinserting this solution into the Lagrangian, using
\be
E^{\pm \mu\nu}_{ij} 
\delta^{ik}\delta^{jl}
E^{\pm \rho \sigma}_{kl} = {}^2\!P^{\mu\nu\rho\sigma}\,,
\ee
we obtain an effective Lagrangian for the metric fluctuations given by 
\be
\mathcal{L}_{\text{eff.}}(h_{\mu\nu}) = A \mathcal{L}_{\text{EH}}(h_{\mu\nu}) - \sum_\pm \frac{\rho_\pm}{4}\qty(\frac{1}{p^2 + M^2_\pm}) p^4 h_{\mu\nu}h_{\rho\sigma} {}^2\!P^{\mu\nu\rho\sigma}\,.
\ee
We can express the last term as
\be
p^4 h_{\mu\nu}h_{\rho\sigma} {}^2\!P^{\mu\nu\rho\sigma} = 2 ^{(1)}\!C_{\mu\nu\rho\sigma}{}^{(1)}\!C^{\mu\nu\rho\sigma}\,,
\ee
where ${}^{(1)}\!C$ denotes the first-order perturbation of the Weyl tensor. Altogether, we obtain an effective metric Lagrangian given by a multiple of the Einstein-Hilbert Lagrangian and a correction quadratic in the Weyl tensor,
\be
\mathcal{L}_{\text{eff.}}(h_{\mu\nu}) = A \mathcal{L}_{\text{EH}}(h_{\mu\nu}) -\frac{1}{2} {}^{(1)}\!C_{\mu\nu\rho\sigma} \qty(\frac{\rho_+}{p^2 + M_+^2} +\frac{\rho_-}{p^2 + M_-^2} ){}^{(1)}\!C^{\mu\nu\rho\sigma}\,.
\ee
The correction term is non-local with the scale of non-locality controlled by the effective mass squares $M_\pm^2 = m_\pm^2/\beta_\pm$. Remarkably, besides $M_\pm^2$, the effective Lagrangian is characterized by only three free parameters, a global rescaling $A$ and the two couplings $\rho_\pm = \alpha_\pm^2/\beta_\pm $. Demanding that we obtain the usual Einstein-Hilbert Lagrangian at leading order in a derivative expansion fixes $A=1$. Next we will show that, if in addition we demand to have no further poles for the spin-2 mode, we obtain the conditions $\rho_++\rho_-=1$ and $M_+=M_-$.  In this case we just remain with one free coupling parameter (corresponding to the Barbero-Immirzi parameter in \cite{BorDitt}) and one mass parameter.

Poles in the effective propagator for the spin-2 mode correspond to zeros of the inverse propagator. Omitting tensorial structures, the latter is given by
\be\label{Prop1}
(\text{Prop}^{-1})^{ \text{spin-2}} = \frac{A}{2}p^2 - \frac{1}{4}\qty(\frac{\rho_+}{p^2 + M_+^2} + \frac{\rho_-}{p^2 + M_-^2})p^4\,.
\ee
Thus we recognize the propagator pole at $p^2=0$ representing the massless spin-2 graviton mode.   If $M_-^2 \neq M_+^2$, we can find additional poles by solving the equation
\be\label{Proppol}
 2 A(p^2 + M_+^2)(p^2 + M_-^2) -             \rho_+\,p^2(p^2 + M_-^2) -\rho_-\,p^2(p^2 + M_+^2) = 0\,.
\ee
 For general couplings $\rho_+,\,\rho_-$ and $A$, the left hand side of this equation yields a second order polynomial in $p^2$. This equation has, in the general case, two solutions for $p^2$, and we thus find additional poles.
 
 If $2A=\rho_++\rho_-$, that is, if (\ref{shiftcond}) holds, the quadratic term in $p^2$ in (\ref{Proppol}) drops out. We obtain a linear equation for $p^2$ which is solved by 
 \ba
 p^2=-M_+^2M_-^2\, \frac{\rho_++\rho_-}{M_+^2\rho_++M_-^2\rho_-} \quad .
 \ea
 
If $M_-^2 = M_+^2=:M^2$ (but $2A\neq\rho_++\rho_-$), additional poles are  again described by an equation linear in $p^2$:
 \ba
 2A(p^2 + M^2) -(\rho_++\rho_-)\,p^2=0 \quad\quad \quad \stackrel{A\neq \rho_++\rho_-}{\Longrightarrow}\quad \quad \quad p^2=-\frac{2A M^2}{2A-\rho_+-\rho_-} \quad .
\ea

Finally, if $2A=\rho_++\rho_-$ and $M_-^2 = M_+^2=:M^2$ holds, the system does not have any additional poles. In this case the propagator is given by
\ba\label{Prop5}
(\text{Prop})^{ \text{spin-2}} = \frac{2}{A}\left( \frac{1}{p^2}+\frac{1}{M^2}\right) \quad ,
\ea
and is independent of how $2A=\rho_++\rho_-$ splits into $\rho_+$ and $\rho_-$. Indeed, Equ.~(\ref{Prop1}) specifies for $\rho_++\rho_-=2A$ and $M_+=M_-=M$  to
\ba
(\text{Prop}^{-1})^{ \text{spin-2}} = \frac{A}{2}p^2 \left(1 -\frac{p^2}{p^2 +M^2}\right) \,=\,  \frac{A}{2}p^2 \frac{M^2}{p^2+M^2} \quad ,
\ea
which inverts to (\ref{Prop5}).  We note that the propagator (\ref{Prop5}) features the same poles as the spin-2 propagator for the gravitons and is therefore ghost-free.

Such a ghost-free propagator has been previously found for two slightly different theories: Firstly in the context of modified chiral Plebanski theory \cite{FreidelMod,Krasnov2}.  The linearized Lagrangian for this theory corresponds to the couplings $\rho_+=A=1,\rho_-=0$ and $M_+\neq 0$ as well as $M_-=0$. The Lagrangian depends therefore only on the fields $h_{\mu\nu}$ and $\chi^+_{ij}$ (and is therefore chiral), and can be written as
\ba
\mathcal{L}_{\text{chiral}}\,=\,\frac{p^2}{2}  \left( h_{\mu\nu} + \, E_{\mu\nu}^{+\, ij} \chi^+_{ij}\right)\big({}^{2}\!P^{\mu\nu\rho\sigma} - 2\,{}^0\! P^{\mu\nu\rho\sigma}\big)  \left( h_{\rho\sigma} + \, E_{\rho\sigma}^{+\, ij} \chi^\pm_{ij}\right)  + M^2 \chi^{ij+}\chi^+_{ij} \quad .
\ea
With the non-local field redefinition (remember that $E_{\mu\nu}^{+\, ij}$ includes inverse derivatives)
\ba
\hat h_{\mu\nu}=h_{\mu\nu} + \, E_{\mu\nu}^{+\, ij} \chi^+_{ij} \; .
\ea
we have
\ba
\mathcal{L}_{\text{chiral}}\,=\,\frac{p^2}{2} \hat h_{\mu\nu}\big({}^{2}\!P^{\mu\nu\rho\sigma} - 2\,{}^0\! P^{\mu\nu\rho\sigma}\big)   \hat h_{\rho\sigma} + M^2 \chi^{ij+}\chi^+_{ij} \quad .
\ea
Here $\chi^{ij+}$ appears only in the mass term and is therefore not propagating. This explains the simple form (\ref{Prop5}) of the propagator for the effective length metric Lagrangian.

The second theory in which the propagator (\ref{Prop5}) appears is modified non-chiral Plebanski theory \cite{Spez,BorDitt}. The linearized theory as discussed in \cite{Spez} features 30 degrees of freedom: two independent length metrics $h^\pm_{\mu\nu}$ and the 10 fields $\chi^\pm_{ij}$. With the aim to construct an effective Lagrangian for spin foams, the paper \cite{BorDitt} argued that spin foam quantization does sharply impose constraints $h^+_{\mu\nu}=h^-_{mu\nu}\equiv h_{\mu\nu}$. This leaves 20 degrees of freedom which can be packaged into area metric perturbations.  The non-chiral Plebanski Lagrangian leads to a linearized Lagrangian which satisfies the condition $1=A=\tfrac{1}{2}(\rho_++\rho_-)${, and the coupling described by the $\rho_\pm$ can be identified with the Barbero-Immirzi parameter. We thus have an additional shift symmetry. But this symmetry only allows for absorbing 5 out of the 10 degrees of freedom encoded in $\chi^+_{ij}$ and $\chi^-_{ij}$. 

One can therefore wonder why the effective length metric Lagrangian features the simple propagator (\ref{Prop5}) also in this case. We will find an answer in the second part of the paper. There (after switching to Lorentzian signature) we will perform a canonical analysis. This will show that we indeed have the usual 2 propagating degrees of freedom of the graviton, and an additional set of 5 degrees of freedom from the $\chi^\pm_{ij}$ fields. Solving the canonical equations of motions will however reveal that there is a variable transformation that allows a decoupling of the dynamics into 2 massless propagating modes and 5 massive propagating modes. This explains the form of the propagator (\ref{Prop5}).

\section{Lorentzian signature and Wick rotation}\label{Sec:WickRotation}

In the first part of the paper we investigated linearized area metric actions in Euclidean signature. We thus used the flat Euclidean  background length metric $(\delta_{\mu\nu}) = \text{diag}(+1,+1,+1,+1)$.  In this second part of the paper we will consider linearized metric actions in Lorentzian signature, that is use the Minkowski metric $(\eta_{\mu\nu})=\text{diag}(-1,+1,+1,+1)$ as a background metric. 

The linearized Einstein-Hilbert action for Minkowskian signature can be straightforwardly obtained from the Euclidean action by contracting indices with the Minkowski metric $\eta$ instead of  the Euclidean metric $\delta$.  This can be also understood as a result of a Wick rotation for the background time coordinate. 

The discussion becomes however more involved for the fields $\chi^\pm_{ij}$ parametrizing our self-dual components. Remember, that these fields arose from the area metric perturbations via
\ba
(\chi^\pm)^{ij}\,=\,    \tfrac{1}{2}    \mathbb{P^\pm}^{ij}_{\mu\nu\rho\sigma} a^{\mu\nu\rho\sigma} \q ,
\ea
where $\mathbb{P^\pm}$ is quadratic in the  Plebanski two-forms ${\Sigma^\pm}^i_{\mu\nu}$ (evaluated on a flat background). In Euclidean signature  these were given by
\ba
{}^{E}\!{\Sigma^\pm}^i_{\mu\nu} &=&  \pm ({\delta}^0_\mu {\delta}^i_\nu-{\delta}^0_\nu {\delta}^i_\mu)+{\epsilon^i}_{jk}{\delta}^j_\mu {\delta}^k_\nu .
\ea
The Kronecker Delta's $\delta^I_\mu$ arise from tetrad variables $e^I_\mu$. Thus if we Wick-rotate, we should multiply $\delta^0_\mu$ with an $\imath$. This leads us to the Lorentzian Plebanski two-forms
\ba\label{Lor2form}
{}^{L}\!{\Sigma^\pm}^i_{\mu\nu} &=&  \pm \imath ({\delta}^0_\mu {\delta}^i_\nu-{\delta}^0_\nu {\delta}^i_\mu)+{\epsilon^i}_{jk}{\delta}^j_\mu {\delta}^k_\nu .
\ea
Indeed, the Lorentzian self-duality condition is an equation with complex coefficients, and for the Minkowski metric given by
\ba
\frac{1}{2}\eta_{\mu\rho}\eta_{\nu\sigma} \epsilon^{\rho\sigma\lambda\tau}\, {}^{L}\!{\Sigma^\pm}^i_{\lambda\tau}\,=\, \pm \imath\, {}^{L}\!{\Sigma^\pm}^i_{\lambda\tau} .
\ea

Assuming a real area metric perturbation, the fields $\chi^\pm_{ij}$ are now complex, but also satisfy 
\ba\label{RealityC1}
\overline{\chi^+_{ij}}=\chi^-_{ij} \q .
\ea
We thus have as many real fields ${\chi_1}_{ij}=\text{Re}(\chi^+_{ij})$ and  ${\chi_2}_{ij}=\text{Im}(\chi^+_{ij})$ as before.  

Now let us consider a term in the Lagrangian involving these complex fields, e.g. 
\ba
{\cal L}_{mass}&=& \mu_+( \chi^+_{ij})^2 + \mu_-( \chi^-_{ij})^2     \q .
\ea
As we have complex fields, we also allow for complex coupling constants. But we demand a real action, which (with (\ref{RealityC1}))  imposes $\overline{\mu_+}=\mu_-$. Introducing a real parametrization for these couplings  $\mu=\mu_1+\imath \mu_2$, we obtain
\ba
{\cal L}_{mass}&=& 2\mu_ 1 ( ({\chi_1}_{ij})^2-({\chi_2}_{ij})^2) -4 \mu_2  {\chi_1}_{ij}   {\chi_2}^{ij}   \q .
\ea

Let us assume that $\mu_+=\mu_-=\mu$ is real and thus $\mu_2=0$. Indeed, we needed to assume equal masses for the plus and minus sector in order to find no additional poles in the length-metric effective action in Section \ref{Sec:EffectiveLengthMetricActions}. (These calculations proceed in the same way for a Minkowskian background.) We then see that whereas we have a positive definite term (for positive $\mu$) in the Lagrangian for an Euclidean background, we also have  terms with indefinite signature for the Lorentzian Lagrangian.

The reader might be already familiar with such an effect for the case of electromagnetism. There one can apply a self-dual decomposition for the electromagnetic field tensor (or two-form) $F_{\mu\nu}$, whose Lagrangian density is proportional to $F_{\mu\nu}F^{\mu\nu}= F^+_{\mu\nu} {F^+}^{\mu\nu}+F^-_{\mu\nu} {F^-}^{\mu\nu}$.  The real and imaginary parts of the self-dual electromagnetic field tensor can be parametrized by the electric and magnetic fields $E^a$ and $B^a$. And here we do also have a similar change in definiteness going from the Euclidean Lagrangian density $\sim E^2+B^2$ to the Lorentzian Lagrangian density $\sim E^2-B^2$.  In this case this change does however not imply any issues for the stability of the Lorentzian system: The $E^2$ term includes the time derivatives of the electro-magnetic potential and is positive definite, whereas $-B^2$ term contains the spatial derivatives, and can be understood as minus the potential energy term, that usually appears in the Lorentzian action. 

Here we have, however, the $\chi^\pm_{ij}$ as fundamental fields, that is they do not arise as derivatives from an underlying field.  We will indeed have an indefinite kinetic term and an indefinite mass term for the $\chi$-fields.  One might therefore expect a non-stable dynamics. To investigate this issue we will perform a canonical analysis of the Lorentzian quadratic area metric action, and consider its evolution. Here we will concentrate on the most interesting case identified in Section \ref{Sec:EffectiveLengthMetricActions}, that is the case with shift symmetry and equal masses. 

We will thus consider the Lorentzian Lagrangian density
\ba
{\cal L}_{full}&=&
A\left(- \frac{1}{2} (\partial_\rho h_{\mu\nu})^2  + (\partial^\nu h_{\mu\nu})^2 - (\partial^\mu h_{\mu\nu}) \partial^\nu h + \frac{1}{2} (\partial_\mu h)^2\right)+ \nn\\
&&
\frac{1}{4}\sum_{\pm}\left(2\alpha_\pm\Sigma^{\pm i}{}^{\mu\nu} \Sigma^{\pm j}{}^{\rho\sigma} h_{\mu\rho} \partial_\nu \partial_\sigma \chi_{ij}^\pm 
-\beta_\pm(\partial_\mu \chi_{ab}^\pm)^2-m^2_\pm ( \chi_{ij}^\pm)^2\right) \q ,
\ea
where from now on $\Sigma^{\pm i}{}^{\mu\nu}=\,{}^L\!\Sigma^{\pm i}{}^{\mu\nu}$.

In the following we will fix the global scale by setting $A=1$. We will also use the freedom to rescale the $\chi^\pm_{ij}$ fields, and in this way achieve $\beta_\pm=1$. We will consider the case $m_+=m_-=m$, as well as $\alpha_+=\overline{\alpha_-}\equiv\alpha_1+\imath \alpha_2$, and later-on restrict to $\alpha_+^2+\alpha_-^2=2\alpha_1^2-2\alpha_2^2=2$.

\section{Hamiltonian analysis}\label{Sec:HamiltonianAnalysis}

In this section we present the Hamiltonian formulation of the area-metric theory in Lorentzian signature. To that end, we first review the constraints and Hamiltonian for linearized general relativity. The linearized Einstein-Hilbert action forms one part of the general  action for the area metric in the parametrization $(h_{\mu\nu}, \chi^\pm_{ij})$. Subsequently, we extend the canonical analysis to the full (quadratic) area-metric action by taking into account the coupling terms between $h_{\mu\nu}$ and $\chi^\pm_{ij}$, as well as the kinetic and mass terms for $\chi^\pm_{ij}$.

\subsection{Constraints and Hamiltonian for linearized general relativity}

The linearized Einstein-Hilbert Lagrangian density is given by
\be
{\cal L}_{\text{EH}} = - \frac{1}{2} (\partial_\rho h_{\mu\nu})^2  + (\partial^\nu h_{\mu\nu})^2 - (\partial^\mu h_{\mu\nu}) \partial^\nu h + \frac{1}{2} (\partial_\mu h)^2\,,
\ee
where $h= h_{\mu\nu}\eta^{\mu\nu} = -h_{00} + h_{ab} \delta^{ab}$ denotes the trace of the metric perturbation. In what follows, $a,b,... = 1,2,3$ are used as spatial indices. To identify the canonically conjugate variables, we perform a $3+1$ decomposition of the Lagrangian. Integrating by parts to remove any time derivatives acting on the time-time and time-space components $h_{00}, h_{0a}$ and dropping all the surface terms, the Lagrangian can be written as

\ba
{\cal L}_{\text{EH}}=  &-&2\partial_b h_{a0}( \dot{h}^{ab}- \delta^{ab} (\delta^{cd}\dot{h}_{cd})) +
\frac{1}{2}(\dot{h}_{ab})^2 - \frac{1}{2}(\dot{h}_{ab}\delta^{ab})^2
+(\partial^a h_{00}) \partial^b (h_{ab} - \delta_{ab}(\delta^{cd}h_{cd}))  \nn\\
&-& (\partial^a h_{a0})^2 
+ (\partial_a h_{b0})^2 
-\frac{1}{2}(\partial_a h_{bc})^2
+ (\partial^b h_{ab})^2 + \delta^{ab}h_{ab} \partial^c \partial^d h_{cd} + \frac{1}{2} (\partial^a \delta^{cd}h_{cd})^2.
\ea

The resulting expression does not include any time derivatives of $h_{00}, h_{a0}$. Thus these components of the metric perturbations represent Lagrange multipliers with vanishing canonical momenta. The only non-vanishing canonical momentum is the one conjugated to the spatial part of the metric, i.e.,
\be\label{eq:p-ab}
p_{ab} \equiv \frac{\partial {\cal L}_{\text{EH}}}{\partial \dot{h}_{ab}} = \dot{h}_{ab}- 2\partial_{(a} h_{b)0} - \delta_{ab} \delta^{cd}(\dot{h}_{cd}- 2\partial_{(c} h_{d)0}).
\ee

Taking the trace of (\ref{eq:p-ab}) allows us to solve for the velocities $\dot{h}_{ab}$ as functions of the momenta $p_{ab}$,
\be\label{eq:hdot}
\dot{h}_{ab}= p_{ab}- \frac{1}{2}\delta_{ab} (\delta^{cd}p_{cd}) + 2\partial_{(a} h_{b)0}\,.
\ee
Additionally integrating by parts the $(\partial_{(a} h_{b)0})^2$ term, we can finally express the Lagrangian (modulo surface terms) in terms of $\dot{h}_{ab}$, its conjugate momentum $p_{ab}$ and the Lagrange multipliers $h_{00},h_{0a}$,
\be
{\cal L}_{\text{EH}}= \frac{1}{2}(p_{ab})^2-\frac{1}{4}(\delta^{ab}p_{ab})^2 +(\partial^a h_{00}) \partial^b (h_{ab} - \delta_{ab}(\delta^{cd}h_{cd}))  
-\frac{1}{2}(\partial_a h_{bc})^2
+ (\partial^b h_{ab})^2 + \delta^{ab}h_{ab} \partial^c \partial^d h_{cd} + \frac{1}{2} (\partial^a \delta^{cd}h_{cd})^2\,.
\ee

The canonical Hamiltonian therefore takes the form
\ba\label{eq:HamiltonianGR}
{\cal H}_{\text{EH}}\equiv p^{ab}\dot{h}_{ab}-{\cal L}_{\text{EH}}&=&
\frac{1}{2}(p_{ab})^2 -\frac{1}{4}(\delta^{ab}p_{ab})^2 
+\frac{1}{2}(\partial_a h_{bc})^2
- (\partial^b h_{ab})^2 - \delta^{ab}h_{ab} \partial^c \partial^d h_{cd} - \frac{1}{2} (\partial^a \delta^{cd}h_{cd})^2
 \nn \\
&+&h_{00} \partial^a\partial^b (h_{ab} - \delta_{ab}(\delta^{cd}h_{cd}))
- 2h_{a0}\partial_b p^{ab}. \nn
\ea

To arrive at the above form of the Hamiltonian for linearized general relativity, we have integrated by parts in the Lagrangian to remove all derivatives from the fields $h_{00}$ and $ h_{a0}$, thereby identifying them as Lagrange multipliers which impose the four first-class primary Hamiltonian and diffeomorphism constraints
\be
C = \partial^a \partial^b (h_{ab}-\delta_{ab}(\delta^{cd}h_{cd}))\,,\quad\quad C_b = \partial^a p_{ab}\,.
\ee
These do not give rise to further secondary constraints. Thus the physical phase space is obtained from the kinematical phase space $(h_{ab}, p_{ab})$ by imposing the four first-class constraints. Each of the four first-class constraints removes two degrees of freedom. The reduced or physical phase space is therefore $2\times 6  - 2\times 4 =  4$-dimensional, describing two propagating degrees of freedom which correspond to the massless spin-2 graviton of general relativity. 

We can explicitly describe the physical degrees of freedom by imposing gauge-fixing conditions 
\be
\delta^{ab} p_{ab}=0\,, \quad\quad \partial^a h_{ab}=0.
\ee
The first condition gauge-fixes the transformations generated by the Hamiltonian constraint, while the second condition gauge-fixes the diffeomorphisms. With this gauge choice the Hamiltonian constraint becomes $\delta (\delta^{ab} h_{ab})=0$, which implies that also $h_{ab}$ is tracefree. In summary, we have two tracefree-transverse ($tt$) fields $(h_{ab}^{tt}, p_{ab}^{tt})$. The physical Hamiltonian is manifestly non-negative and given by
\be
{\cal H}^{phys}_{\text{EH}} = \frac{1}{2}(p_{ab}^{tt})^2 + \frac{1}{2} (\partial_c h_{ab}^{tt})^2.
\ee

\subsection{$3+1$ decomposition of the Lorentzian area-metric action}

We will now consider the Lagrangian density for the $h_{\mu\nu}$ and the $\chi^\pm_{ij}$ fluctuations, given by
\ba
{\cal L}_{full}&=&
{\cal L}_{\text{EH}}+
\frac{1}{4}\sum_{\pm}\left(2\alpha_\pm\Sigma^{\pm a}{}^{\mu\nu} \Sigma^{\pm b}{}^{\rho\sigma} h_{\mu\rho} \partial_\nu \partial_\sigma \chi_{ab}^\pm 
-(\partial_\mu \chi_{ab}^\pm)^2-m^2 (\chi_{ab}^\pm)^2\right) \q ,
\ea
Here we identified the internal indices $i,j,k,\ldots$ with spatial indices $a,b,c,\ldots$. This is possible via the background spatial triads $e^i_a=\delta^i_a$. Remember that $\chi^\pm=\chi_1\pm\imath \chi_2$ and $\alpha_\pm=\alpha_1\pm\imath \alpha_2$.

The most involved new term in this Lagrangian is the term describing the coupling between the $\chi$ and $h$-fields.

Using the explicit expression~\eqref{Lor2form} for the Plebanski two-forms, this term can be written after partial integration as 
\ba
&&\tfrac{1}{2}\sum_{\pm}\alpha_\pm\Sigma^{\pm a}{}^{\mu\nu} \Sigma^{\pm b}{}^{\rho\sigma} h_{\mu\rho} \partial_\nu \partial_\sigma \chi_{ab}^\pm 
\nn\\
&=& \sum_{1,2} \pm \alpha_{1,2}\Big[-h_{00}\partial_a \partial_b \chi^{ab}_{1,2} - 2 \partial_a h_{b0} \dot{\chi}^{ab}_{1,2} + \dot{h}_{ab}\dot{\chi}^{ab}_{1,2} 
+ \epsilon^{aef}\epsilon^{bpq} h_{ep}\partial_f \partial_q \chi_{1,2\,ab}\Big]
+2\alpha_{1,2}( \dot{h}^{ab} -2\partial^{(a} h^{b)0})
\epsilon_a{}^{ef}\partial_e \chi_{2,1\, fb}\,.\nn
\ea
We can write the previous expression more compactly by introducing the combination
\be
\chi_{ab} \equiv \alpha_1 \chi_{1\, ab} - \alpha_2 \chi_{2\, ab}.
\ee
and recalling the relation (\ref{eq:p-ab}) between the momentum $p_{ab}$  and $\dot{h}_{ab}$ (which are conjugated variables in linearized general relativity). Herewith we can rewrite the coupling term as
\be
\tfrac{1}{2}\sum_{\pm}\alpha_\pm\Sigma^{\pm a}{}^{\mu\nu} \Sigma^{\pm b}{}^{\rho\sigma} h_{\mu\rho} \partial_\nu \partial_\sigma \chi_{ab}^\pm  = -h_{00}\partial_a \partial_b \chi^{ab}  + p_{ab} \dot{\chi}^{ab} 
+ \epsilon^{aef}\epsilon^{bpq} h_{ep}\partial_f \partial_q \chi_{ab}
+2p^{ab}
\epsilon_a{}^{ef}\partial_e (\alpha_1 \chi_{2\, fb}+\alpha_2 \chi_{1\, fb}),
\ee
where we used that $\chi_{1,2}$, as well as $\epsilon_{(a}{}^{ef}\partial_{|e} \chi_{f|b)}$, are tracefree. The full Lagrangian can thus be written as
\ba\label{L-full}
{\cal L}_{full}(h_{\mu\nu},\chi_{1,2\,ab})  &=&
\frac{1}{2}(p_{ab})^2-\frac{1}{4}(\delta^{ab}p_{ab})^2 -h_{00}\partial^a \partial^b (h_{ab} - \delta_{ab}(\delta^{cd}h_{cd}))  
\nn\\ 
&-&\frac{1}{2}(\partial_a h_{bc})^2
+ (\partial^b h_{ab})^2 + \delta^{ab}h_{ab} \partial^c \partial^d h_{cd} + \frac{1}{2} (\partial^a \delta^{cd}h_{cd})^2
\nn\\ 
&-&h_{00}\partial_a \partial_b \chi^{ab}  + p_{ab} \dot{\chi}^{ab} 
+ \epsilon^{aef}\epsilon^{bpq} h_{ep}\partial_f \partial_q \chi_{ab}
+2p^{ab}
\epsilon_a{}^{ef}\partial_e (\alpha_1 \chi_{2\, fb}+\alpha_2 \chi_{1\, fb})
\nn\\ 
&+&\frac{1}{2}((\dot{\chi}_{1\, ab})^2- (\dot{\chi}_{2\, ab})^2) - \frac{1}{2}( (\partial_a\chi_{1\, bc})^2 - (\partial_a \chi_{2\, bc})^2)
\,\,-\,\, \frac{1}{2}m^2  \left((\chi_1^{ab})^2 - (\chi_2^{ab})^2\right)\,.
\ea

The conjugate momenta are determined by
\ba\label{momenta-h-chi}
P_{ab} &\equiv & \frac{\partial {\cal L}_{full}}{\partial \dot{h}_{ab}} = p_{ab}+ \alpha_1 \dot{\chi}_{1\, ab}
- \alpha_2 \dot{\chi}_{2\, ab} + 2\epsilon_{(a}{}^{ef}\partial_{|e} (2\alpha_2 \chi_{1\, f|b)}+2\alpha_1 \chi_{2\, f|b)})\,,
\\ \nonumber
\rho_{1\, ab} &\equiv & \frac{\partial {\cal L}_{full}}{\partial \dot{\chi}_{1\, ab}} = \dot{\chi}_{1\, ab} + \alpha_1 p_{ab} - \frac{1}{3} \alpha_1\delta_{ab}
\delta^{cd}p_{cd}\,,
\\ \nonumber
\rho_{2\, ab} &\equiv & \frac{\partial {\cal L}_{full}}{\partial \dot{\chi}_{2\, ab}} = -\dot{\chi}_{2\, ab} -\alpha_2 p_{ab} + \frac{1}{3} \alpha_2\delta_{ab}
\delta^{cd}p_{cd}\,.
\ea
Here we have explicitly subtracted the traces to make $\rho_{1,2}$ tracefree. 

As a next step, the velocities have to be solved as functions of the momenta. The trace part of the first equation is insensitive to $\chi_{1,2}$ as these fields are tracefree, and thus it can be solved in the same way as for linearized general relativity. In particular, only the tracefree parts of the equations for the conjugate momenta are coupled. This system is non-singular unless
\be\label{relation}
\alpha_1^2 - \alpha_2^2 =1.
\ee
If this condition is satisfied, the rank of the system of equations (in the space of three symmetric tracefree fields) is two rather than three, which signalizes the presence of the additional "shift" symmetry. In the following we will restrict the analysis to the case when~\eqref{relation} is satisfied. To that end, in the next subsection we introduce an alternative basis of fields to parametrize the Lagrangian.

\subsection{Reparametrization of the Lagrangian}

The relation~\eqref{relation} implies that we can parametrize the real and imaginary parts of the complex parameter $\alpha_+=\alpha_1+\imath \alpha_2$ in terms of a single real parameter $\xi$ by writing~\footnote{Note that one can use a field redefinition $\chi^\pm\rightarrow -\chi^\pm$ to change a negative coupling $\alpha_1$ into a positive coupling.}
\be
\alpha_1 = \cosh(\xi)\,, \qquad \alpha_2 = \sinh(\xi).
\ee
Introducing the field redefinitions
\be
\chi \equiv  \cosh(\xi)\chi_1 - \sinh(\xi) \chi_2\,, \qquad \phi \equiv \sinh(\xi) \chi_1 - \cosh(\xi) \chi_2
\ee
allows us to express the full Lagrangian~\eqref{L-full} after some algebraic manipulations in the form
\ba
{\cal L}_{full}(h_{\mu\nu},\chi_{ab},\phi_{ab}) &=& 
\frac{1}{2}(p_{ab}+\dot{\chi}_{ab})^2-\frac{1}{4}(\delta^{ab}(p_{ab}+\dot{\chi}_{ab}))^2 -h_{00}\partial^a \partial^b (h_{ab} +\chi_{ab} - \delta_{ab}(\delta^{cd}(h_{cd}+\chi_{cd}))  
\nn\\ 
&-&\frac{1}{2}(\partial_a h_{bc})^2
+ (\partial^b h_{ab})^2 + \delta^{ab}h_{ab} \partial^c \partial^d h_{cd} + \frac{1}{2} (\partial^a \delta^{cd}h_{cd})^2
\nn\\ 
&+& \epsilon^{aef}\epsilon^{bpq} h_{ep}\partial_f \partial_q \chi_{ab}
+2p^{ab}
\epsilon_a{}^{ef}\partial_e (\sinh(2\xi) \chi_{fb} - \cosh(2\xi) \phi_{fb})
\nn\\ 
&-&\frac{1}{2} (\dot{\phi}_{ab})^2 - \frac{1}{2}( (\partial_a\chi_{bc})^2 - (\partial_a \phi_{bc})^2) - \frac{1}{2}m^2 (\chi_{ab})^2 + \frac{1}{2}m^2 (\phi_{ab})^2 .
\ea
From the above expression we recognize that the terms $p^2, p\dot{\chi}$ and $\dot{\chi}^2$ form a perfect square. As a next step we rewrite the Lagrangian in terms of 
\ba\label{eq:PPhi}
P_{ab}& \equiv & p_{ab}+\dot{\chi}_{ab}+   2 \epsilon_{(a}{}^{ef}\partial_{|e} (\sinh(2\xi) \chi_{f|b)} - \cosh(2\xi) \phi_{f|b)})\,,\nn\\
\Phi_{ab} &\equiv & -\dot{\phi}_{ab}- 2\cosh(2\xi) \epsilon_{(a}{}^{ef}\partial_{|e}  \chi_{f|b)}\,.
\ea
We can use partial integration repeatedly to see that the term $\chi\dot{\chi}$ vanishes (modulo a surface term), whereas the term $\dot{\chi}\phi$ can be rewritten such that the time derivative acts on $\phi$. Herewith the Lagrangian can be expressed as 
\ba\label{eq:L-full-1}
{\cal L}_{full}(h_{\mu\nu},\chi_{ab},\phi_{ab}) &=& 
\frac{1}{2}(P_{ab})^2-\frac{1}{4}(\delta^{ab}P_{ab})^2 
-h_{00}\partial^a \partial^b (h_{ab} +\chi_{ab} - \delta_{ab}(\delta^{cd}(h_{cd}+\chi_{cd}))  
\nn\\ 
&-&\frac{1}{2}(\partial_a h_{bc})^2
+ (\partial^b h_{ab})^2 + \delta^{ab}h_{ab} \partial^c \partial^d h_{cd} + \frac{1}{2} (\partial^a \delta^{cd}h_{cd})^2
\nn\\ 
&+& \epsilon^{aef}\epsilon^{bpq} h_{ep}\partial_f \partial_q \chi_{ab}
-2(\epsilon_{(a}{}^{ef}\partial_{|e} (\sinh(2\xi) \chi_{f|b)} - \cosh(2\xi) \phi_{f|b)}))^2 + 2 \cosh^2(2\xi) (\epsilon_{(a}{}^{ef}\partial_{|e} (\chi_{f|b)} )^2
\nn\\ 
&-& \frac{1}{2}(\Phi_{ab})^2-\frac{1}{2}( (\partial_a\chi_{bc})^2 - (\partial_a \phi_{bc})^2) - \frac{1}{2}m^2 (\chi_{ab})^2 + \frac{1}{2}m^2(\phi_{ab})^2.
\ea

From~\eqref{eq:L-full-1} we see that the spatial metric subject to the Hamiltonian constraint is
\be\label{H}
H_{ab} \equiv h_{ab}+ \chi_{ab}.
\ee
Rewriting all potential terms in terms of this redefined metric, using partial integration, we can write down the full Lagrangian in the form
\ba\label{L-full-final}
{\cal L}_{full}(H_{ab},\chi_{ab},\phi_{ab}) &=& 
\frac{1}{2}(P_{ab})^2-\frac{1}{4}(\delta^{ab}P_{ab})^2 
-h_{00}\partial^a \partial^b (H_{ab} - \delta_{ab}(\delta^{cd}H_{cd}))  
\nn\\ 
&-&\frac{1}{2}(\partial_a H_{bc})^2
+ (\partial^b H_{ab})^2 + \delta^{ab}H_{ab} \partial^c \partial^d H_{cd} + \frac{1}{2} (\partial^a \delta^{cd}H_{cd})^2
\nn\\ 
&+& 2\epsilon^{aef}\epsilon^{bpq} H_{ep}\partial_f \partial_q \chi_{ab}+ 4 \cosh(2\xi)\sinh(2\xi) (\epsilon_{(a}{}^{ef}\partial_{|e} \chi_{f|b)} )(\epsilon^a{}_{cd}\partial^c \phi^{db} )- \frac{1}{2}m^2 (\chi_{ab})^2 
\nn\\ 
&-&\frac{1}{2}(\Phi_{ab})^2 + \frac{1}{2}(\partial_a \phi_{bc})^2 - 2\cosh^2(2\xi) (\epsilon_{(a}{}^{ef}\partial_{|e} \phi_{f|b)} )^2 + \frac{1}{2}m^2(\phi_{ab})^2\,.
\ea
The first two lines represent the linearised Einstein-Hilbert Lagrangian for the spatial metric $H_{ab}$ and its conjugated momentum $P_{ab}$ (as can be easily checked) with the Lagrange multiplier $h_{00}$ imposing the Hamiltonian constraint and $h_{0a}$ imposing the diffeomorphism constraint (as can be seen by inserting the explicit expression for $P_{ab}$). The last line is a Lagrangian for the field $\phi_{ab}$ which has notably a wrong sign in front of its kinetic and mass term. The third line involves the field $\chi_{ab}$, whose momentum is constrained to vanish.

To express the Lagrangian in terms of the fields $H_{ab}, \chi_{ab}, \phi_{ab}$ and their time derivatives we use that, with~\eqref{eq:PPhi} and~\eqref{eq:p-ab} we have
\ba\label{momenta2}
P_{ab}&=&\dot{H}_{ab}-
\delta_{ab} \delta^{cd}\dot{H}_{cd} 
-2\partial_{(a} h_{b)0} + 2 \delta_{ab} \delta^{cd} \partial_c h_{d0} +
2 \sinh(2 \xi) (D\chi)_{ab} - 2 \cosh(2\xi) (D\phi)_{ab}\,, \nn\\
\Phi_{ab}&=& -\dot{\phi}_{ab}-2\cosh(2\xi) (D\chi)_{ab}
\ea
where $Dt_{ab}\equiv\epsilon_{(a}{}^{ef}\partial_{|e} t_{f|b)}$.   Thus, the Lagrangian density expressed as a functional of $H_{ab},\chi_{ab}$ and $\phi_{ab}$ is given by
\ba\label{L-mass}
{\cal L}_{full}(H_{ab},\chi_{ab},\phi_{ab}) &=& 
+\frac{1}{2}\left(
\dot{H}_{ab}-
\delta_{ab} \delta^{cd}\dot{H}_{cd} 
-2\partial_{(a} h_{b)0} + 2 \delta_{ab} \delta^{cd} \partial_c h_{d0} +
2 \sinh(2 \xi) (D\chi)_{ab} - 2 \cosh(2\xi) (D\phi)_{ab}
\right)^2 \nn\\
&&-\frac{1}{4}\left( -2 \delta^{ab} \dot H_{ab} +4 \delta^{ab} \partial_a h_{b0}   \right)^2 
-h_{00}\partial^a \partial^b (H_{ab} - \delta_{ab}(\delta^{cd}H_{cd}))  
\nn\\
&&-\frac{1}{2}(\partial_a H_{bc})^2
+ (\partial^b H_{ab})^2 + \delta^{ab}H_{ab} \partial^c \partial^d H_{cd} + \frac{1}{2} (\partial^a \delta^{cd}H_{cd})^2
 \nn\\
&&+ 2\epsilon^{aef}\epsilon^{bpq} H_{ep}\partial_f \partial_q \chi_{ab}+ 4 \cosh(2\xi)\sinh(2\xi) (D\chi)_{ab}(D\phi )^{ab} -\frac{1}{2}m^2(\chi_{ab})^2
 \nn\\
&&-\frac{1}{2}( \dot{\phi}_{ab} + 2 \cosh(2 \xi) (D \chi)_{ab}    )^2
+ \frac{1}{2}(\partial_a \phi_{bc})^2 - 2\cosh^2(2\xi) ((D\phi)_{ab} )^2 + \frac{1}{2}m^2(\phi_{ab})^2\,.
\ea

\subsection{Hamiltonian}

Only the fields $H_{ab}$
and $\phi_{ab}$ have non-vanishing conjugate momenta equal to the already introduced combinations $P_{ab},\Phi_{ab}$,
\be
\frac{\partial {\cal L}_{full}}{\partial \dot{H}_{ab}} = P^{ab}, \qquad
\frac{\partial {\cal L}_{full}}{\partial \dot{\phi}_{ab}} = \Phi^{ab}.
\ee
To obtain the Hamiltonian, we need to compute $P^{ab}\dot{H}_{ab}$ and $\Phi^{ab}\dot{\phi}_{ab}$. These are derived using 
\ba\label{eq: Hdotphidot}
\dot{H}_{ab} &=& P_{ab} + 2\partial_{(a}h_{b)0} - \frac{1}{2}\delta_{ab} \delta^{cd}P_{cd} - 2\epsilon_{(a}{}^{ef}\partial_{|e} (\sinh(2\xi) \chi_{f|b)} - \cosh(2\xi) \phi_{f|b)}), 
\nn\\
\dot{\phi}_{ab} &=& -\Phi_{ab} - 2\cosh(2\xi) \epsilon_{(a}{}^{ef}\partial_{|e}  \chi_{f|b)}.
\ea
Introducing the differential operators $Dt_{ab}=\epsilon_{(a}{}^{ef}\partial_{|e} t_{f|b)}$ 
and $({\mathbb D} t)_{ab}= {\epsilon_a}^{fc} {\epsilon_b}^{qd}\partial_f \partial_q t_{cd}$
we can now write down the Hamiltonian density  (using integrations by part)
\ba\label{H-full}
{\cal H}_{full}&=&P^{ab}\dot{H}_{ab}+ \Phi^{ab}\dot{\phi}_{ab} - {\cal L}_{full}\nn\\
&\hat{=}&
\frac{1}{2}(P_{ab})^2-\frac{1}{4}(\delta^{ab}P_{ab})^2 
+h_{00}\partial^a \partial^b (H_{ab} - \delta_{ab}(\delta^{cd}H_{cd}))  + 2P^{ab}\partial_a h_{b0}
\\ \nn
&&+\frac{1}{2}(\partial_a H_{bc})^2
- (\partial^b H_{ab})^2 - \delta^{ab}H_{ab} \partial^c \partial^d H_{cd} - \frac{1}{2} (\partial^a \delta^{cd}H_{cd})^2
\\ \nn
&&- 2P^{ab} (\sinh(2\xi) D\chi_{ab} - \cosh(2\xi) D\phi_{ab})- 2\cosh(2\xi) \Phi^{ab} D\chi_{ab}
\\ \nn
&&- 2 \chi^{ab} {\mathbb D} H_{ab} - 4 \cosh(2\xi)\sinh(2\xi) D \chi_{ab} D \phi^{ab} 
\\ \nn
&&-\frac{1}{2}(\Phi_{ab})^2 - \frac{1}{2}(\partial_a \phi_{bc})^2 + 2\cosh^2(2\xi) (D\phi_{ab} )^2 + \frac{1}{2} m^2 (\chi_{ab})^2 -\frac{1}{2} m^2 (\phi_{ab})^2 \quad .
\ea
Note that the first two lines of the Hamiltonian density agree with the Hamiltonian density for linearized GR, see equation~\eqref{eq:HamiltonianGR}, if we replace $h_{ab}\rightarrow H_{ab}$ and $p_{ab}\rightarrow P_{ab}$.

This system  has the following constraints: Firstly, we have the usual linearized constraints of general relativity 
\ba
C=\partial^a \partial^b (H_{ab} - \delta_{ab}(\delta^{cd}H_{cd})) \quad, \quad\quad
C_b=\partial^a P_{ab}     \quad .
\ea
These commute with each other. But we also have to check the time evolution of these constraints with the Hamiltonian.  It turns out that the constraints commute with the part of the Hamiltonian involving the $\chi$ and $\phi$ fields and their conjugated momenta. Thus one has the same commutation relations as in linearized GR, namely
\ba\label{commC}
\{C,\int d^3x \, {\cal H}_{full} \} = \partial^b C_b \,\quad \quad  \{C_b,\int d^3x \,{\cal H}_{full} \}=0 \quad. 
\ea
We thus do not generate any further constraints. 

Additionally we have the primary constraints that the momenta $\Theta_{ab}$ conjugated to $\chi_{ab}$ vanish: $\Theta_{ab}=0$. We have to consider also time evolution of these primary constraints. To this end, note that for  symmetric tensors $s^{ab}$ and $t_{ab}$, using integration by parts, we have $(Ds)^{ab}t_{ab}\hat{=} s^{ab} (Dt)_{ab}$. We thus obtain the following secondary constraints:
\ba
C_{ab}
&=& 
m^2\chi_{ab}-2\sinh(2\xi)DP_{ab}-2 \cosh(2\xi) D\Phi_{ab} - 2 {\mathbb D} H_{ab} - 4\cosh(2\xi) \sinh(2\xi) D^2 \phi_{ab}   \nn\\
&=:& m^2\chi_{ab}-F_{ab}(H_{cd},P_{cd},\phi_{cd},\Phi_{cd}) \,.
\ea

The primary and secondary constraints form a second-class system. (The commutator of $F_{ab}$ with $F_{cd}$ vanishes and therefore $C_{ab}$ commutes with  $C_{cd}$.) Following Dirac's procedure we have to add the primary and secondary constraints multiplied with Lagrange multipliers to the system. Demanding that the constraints are preserved by time evolution fixes these Lagrange multipliers, and we thus do not generate further constraints. 

We can eliminate the constraints $\Theta_{ab}$ and $C_{ab}$ from the system by solving the $C_{ab}$ for $\chi_{ab}$ and by inserting this solution into the Hamiltonian density. 

$C_{ab}$ and $\Theta_{ab}$ are conjugated to each other, the Dirac brackets amount therefore to the Poisson brackets, if we restrict to the variables $H,P$ and $\phi,\Phi$. 

We  are thus left with a reduced Hamiltonian density
\ba\label{Hprime}
{\cal H}'_{full}
&=&
\frac{1}{2}\left(P_{ab}+ 2 \cosh(2\xi) D\phi_{ab}\right)^2-\frac{1}{4}(\delta^{ab}P_{ab})^2 
+h_{00}\partial^a \partial^b (H_{ab} - \delta_{ab}(\delta^{cd}H_{cd}))  + 2P^{ab}\partial_a h_{b0}
 \nn\\
&&+\frac{1}{2}(\partial_a H_{bc})^2
- (\partial^b H_{ab})^2 - \delta^{ab}H_{ab} \partial^c \partial^d H_{cd} - \frac{1}{2} (\partial^a \delta^{cd}H_{cd})^2
\nn\\ 
&&-\frac{1}{2}(\Phi_{ab})^2 - \frac{1}{2}(\partial_a \phi_{bc})^2  - \frac{1}{2m^2}  (F_{ab})^2 -\frac{1}{2} m^2 (\phi_{ab})^2 \quad .
\ea

Here we notice that all terms in the last line of (\ref{Hprime}) are negative definite. One might thus indeed fear that the dynamics will be unstable. To investigate this issue we will solve the time evolution equations.

The time evolution equations are given by:
%
%
\ba\label{eq:HamiltonianEOM}
\dot{H}_{ab} =\{H_{ab},\int d^3\!x\, {\cal H}'_{full} \} 
&=& P_{ab} + 2\cosh(2\xi) D\phi_{ab} -\frac{1}{2}\delta_{ab} \delta^{cd}P_{cd} + 2 \delta^c_{(a} \delta^d_{b)} \partial_c h_{d0}     - \frac{2}{m^2}\sinh(2\xi) DF_{ab} \, ,
\nn\\
\dot{P}_{ab} = \{P_{ab},\int d^3\!x\, {\cal H}'_{full} \} 
&=& \partial_c\partial^c H_{ab} 
-\partial_a\partial_b h_{00} + \delta_{ab}\partial^c\partial_c h_{00} \nn\\
&&-\partial_a\partial^cH_{cb}-\partial_b\partial^cH_{ca} + \delta_{ab}\partial^c\partial^dH_{cd} +\partial_a\partial_b \delta^{cd}H_{cd}-\delta_{ab}\partial^e\partial_e \delta^{cd} H_{cd}
+\frac{2}{m^2} {\mathbb D}   F_{ab}
\,,\nn\\
\dot{\phi}_{ab} =  \{\phi_{ab},\int d^3\!x\, {\cal H}'_{full} \} 
&=& -\Phi_{ab} -\frac{2}{m^2}\cosh(2\xi)DF_{ab} \, ,\nn\\
\dot{\Phi}_{ab} = \{\Phi_{ab},\int d^3\!x\, {\cal H}'_{full} \} 
&=& (-\partial_c\partial^c- 4\cosh(2\xi)^2D^2 +m^2)\phi_{ab} - 2\cosh(2\xi)DP_{ab} +\frac{4}{m^2} \cosh(2\xi)\sinh(2\xi)D^2F_{ab}\,, \nn\\
\ea
where
\ba
F_{ab}&=&
2\sinh(2\xi)DP_{ab}+2 \cosh(2\xi) D\Phi_{ab} + 2 {\mathbb D} H_{ab} + 4\cosh(2\xi) \sinh(2\xi) D^2 \phi_{ab}  \quad .
\ea

\subsection{Mode decomposition and differential operators}

In order to solve these differential equations we apply a Fourier transform in the spatial coordinates, which replaces $\partial_a \rightarrow \imath k_a$. We furthermore use a  decomposition of the symmetric tensor modes as follows. 
For the transverse-traceless modes we use the orthonormal basis given by
\be\label{PlusCross}
t_{ab}^+ \equiv \frac{1}{\sqrt{2}} \qty(\hat{k}_a^\theta \hat{k}_b^\theta - \hat{k}_a^\varphi \hat{k}_b^\varphi )\,,\quad\quad \quad t_{ab}^{\cross} \equiv \frac{1}{\sqrt{2}} \qty(\hat{k}_a^\theta \hat{k}_b^\varphi + \hat{k}_a^\varphi \hat{k}_b^\theta)\,,
\ee
and the alternative (complex) set which diagonalizes helicity
\be
t_{ab}^R \equiv  \frac{1}{2}\left( \hat k^\theta_a+\imath \hat k^\varphi_a\right)\left( \hat k^\theta_b+\imath \hat k^\varphi_b\right)
\,,\quad\quad \quad t_{ab}^L \equiv  \frac{1}{2}\left( \hat k^\theta_a-\imath \hat k_a^\varphi\right)\left( \hat k^\theta_b-\imath \hat k_b^\varphi\right)\,.
\ee
Here the $3$-vectors  $\hat{\vec{k}}^\theta$ and $\hat{\vec{k}}^\varphi$ are such that
$\qty(\frac{\vec{k}}{|\vec{k}|}, \hat{\vec{k}}^\theta,\hat{\vec{k}}^\varphi)$ defines  right-handed orthonormal basis, i.e.,
\be\label{eq:kIdentities}
\hat{k}_a^\theta  \hat{k}_b^\theta \delta^{ab} = \hat{k}_a^\varphi \hat{k}_b^\varphi\delta^{ab} =1\,,\quad\quad\quad \hat{k}_a^\theta  \hat{k}_b^\varphi \delta^{ab} = \hat{k}_a^\theta  k_b \delta^{ab} =\hat{k}_a^\varphi k_b \delta^{ab}  =0\,,\quad\quad\quad \epsilon^{abc}\frac{k_{a}}{|\vec{k}|}\hat{k}_b^\theta \hat{k}_c^\varphi = +1\,.
\ee
We furthermore define the following basis of longitudinal modes, projected to be traceless:
\be\label{eq:DActionLT}
t^{l\theta}_{ab} \equiv \frac{1}{\sqrt{2|\vec{k}|^2}}\qty(k_a \hat{k}_b^\theta + \hat{k}_a^\theta k_b)\,,\quad\quad
t^{l\varphi}_{ab} \equiv \frac{1}{\sqrt{2|\vec{k}|^2}}\qty(k_a \hat{k}_b^\varphi + \hat{k}_a^\varphi k_b)\,,\quad\quad t^{ll}_{ab} \equiv \sqrt{\frac{3}{2|\vec{k}|^4}}\qty(k_ak_b - \frac{1}{3}|\vec{k}|^2 \delta_{ab})\,.
\ee
Note that these modes are orthogonal to the transverse-traceless modes, and that they form in itself an orthonormal basis. 

And alternative pair of (complex) basis vectors, replacing the first two vectors in (\ref{eq:DActionLT}) is
\be
t^{l+}_{ab} \equiv \frac{1}{\sqrt{2}}(t^{l\theta}_{ab}+\imath t^{l\varphi}_{ab})
\,,\quad\quad
t^{l-}_{ab} \equiv \frac{1}{\sqrt{2}}(t^{l\theta}_{ab}-\imath t^{l\varphi}_{ab})\,.
\ee

Finally we have a trace mode. We choose this mode to be orthogonal to all the modes listed above and normalized to one:
\be
t^{tr}_{ab}=\frac{1}{\sqrt{3}}\delta_{ab} \quad .
\ee

Let us now consider how the operator $D$, defined by $(Dt)_{ab}=\epsilon_{(a}{}^{ef}\partial_{|e} t_{f|b)}$ acts on these modes. It is straightforward to compute 
\begin{align}
 (Dt^+)_{ab}&=\imath |\vec{k}| t^{\cross}_{ab} \,  ,&
(Dt^{\cross})_{ab}&=-\imath |\vec{k}| t^+_{ab} \,,\nn\\
 (Dt^R)_{ab}&= |\vec{k}| t^R_{ab} \,  ,&
(Dt^L)_{ab}&=- |\vec{k}| t^L_{ab} \,,\nn\\
 (Dt^{l\theta})_{ab}&=\frac{\imath}{2} |\vec{k}| t^{l\varphi}_{ab} \, ,&
 (Dt^{l\varphi})_{ab}&=-\frac{\imath}{2} |\vec{k}| t^{l\theta}_{ab}\,, \nn\\
 (Dt^{l+})_{ab}&= \frac{1}{2} |\vec{k}| t^{l+}_{ab} \, ,&
 (Dt^{l-})_{ab}&=-\frac{1}{2} |\vec{k}| t^{l-}_{ab}\,, \nn\\
(Dt^{ll})_{ab}&=0 \,  , &
(Dt^{tr})_{ab}&=0 \, \, .
\end{align}

Note  that for the transverse-traceless modes $t^{tt}_{ab}$ we have
\ba
(D^2 t^{tt})_{ab}\,=\,-\partial^c\partial_c t^{tt}_{ab} \,=\, |\vec{k}|^2  t^{tt}_{ab} \quad .
\ea
That is $D$ is a square root of minus the spatial Laplacian, if we restrict to transverse-traceless modes.  
We see that $D$ acts diagonally on the basis elements $t^R,t^L$ and $t^{l+},t^{l-}$, 
but not on the basis elements 
$t^+,t^{\cross}$ 
and $t^{l\theta},t^{l\phi}$.

We have furthermore the second-order differential operator $({\mathbb D} t)_{ab}= {\epsilon_a}^{fc} {\epsilon_b}^{qd}\partial_f \partial_q t_{cd}$. It acts as follows on the various modes:
\begin{align}\label{DDaction}
 ({\mathbb D}t^+)_{ab}&=|\vec{k}|^2 t^{+}_{ab} \,  ,&
({\mathbb D}t^{\cross})_{ab}&=|\vec{k}|^2 t^{\cross}_{ab} \,,\nn\\
({\mathbb D}t^R)_{ab}&=|\vec{k}|^2 t^{R}_{ab} \,  ,&
({\mathbb D}t^L)_{ab}&=|\vec{k}|^2 t^{L}_{ab} \,,\nn\\
 ({\mathbb D}t^{l\theta})_{ab}&=0\, ,&
 ({\mathbb D}t^{l\varphi})_{ab}&=0\,, \nn\\
  ({\mathbb D}t^{l+})_{ab}&=0\, ,&
 ({\mathbb D}t^{l-})_{ab}&=0\,, \nn\\
({\mathbb D}t^{ll})_{ab}&=-\frac{1}{3}|\vec{k}|^2 t^{ll}_{ab}+\frac{\sqrt{2}}{3}|\vec{k}|^2  t^{tr}_{ab} \,  , &
({\mathbb D}t^{tr})_{ab}&=\frac{\sqrt{2}}{3}|\vec{k}|^2 t^{ll}_{ab}-\frac{2}{3}|\vec{k}|^2  t^{tr}_{ab} \, \, .
\end{align}
Note that for the transverse-traceless modes $t^{tt}_{ab}$ we have
\ba
({\mathbb D}t^{tt})_{ab}\,=\,(D^2 t^{tt})_{ab}\,=\, |\vec{k}|^2  t^{tt}_{ab} \quad .
\ea

Introducing an alternative orthonormal basis for the space spanned by the $(ll)$ and $(tr)$ modes 
\ba
t^{dl}_{ab}:=\frac{k_ak_b}{|\vec{k}|^2}=\sqrt{\frac{2}{3}}\left(t^{ll}_{ab}+\frac{1}{\sqrt{2}} t^{tr}_{ab} \right) \quad, \quad \quad
t^t_{ab}:=\frac{1}{\sqrt{2}}\left( \frac{k_ak_b}{|\vec{k}|^2}-\delta_{ab}\right) =
\sqrt{\frac{2}{3}}\left(\frac{1}{\sqrt{2}} t^{ll}_{ab}-t^{tr}_{ab} \right)
\ea
the action of ${\mathbb D}$ simplifies to
\ba
({\mathbb D} t^{dl})_{ab}=0 \, , \quad \quad ({\mathbb D} t^{t})_{ab}=-|\vec{k}|^2 t^t_{ab} \quad .
\ea

\subsection{Solution to the Hamiltonian equations of motion}\label{Sec:SolH}

To solve the Hamiltonian equations of motion~\eqref{eq:HamiltonianEOM} for the dynamical variables $(H_{ab},P_{ab},\phi_{ab},\Phi_{ab})$, we decompose the symmetric tensors into modes, e.g. 
\be
H_{ab} =H^R t_{ab}^R +H^{L} t^{L}_{ab} + H^{l+}t^{l+}_{ab}+ H^{l-}t^{l-}_{ab}+ H^{ll}t^{ll}_{ab}+ H^{tr}t^{tr}_{ab}
\ee
and so on. (Note that $\phi^{tr}=0$ and $\Phi^{tr}=0$.)
It is easy to see that the various differential operators appearing in the equations of motion (\ref{eq:HamiltonianEOM}) act diagonally on the basis elements labelled by $\{R,L,l+,l-\}$, but not on $\{ll,tr\}$. 
We therefore obtain a closed subsystem for each label in the first set.

For the $R/L$ modes we can write
\ba
\qty(\begin{matrix}
	\dot{H}^{R/L}  \\
	\dot{P}^{R/L}   \\
		\dot{\phi}^{R/L} \\
				\dot{\Phi}^{R/L} 
\end{matrix}) 
&=&
M^{R/L} \, \cdot\,
\qty(\begin{matrix}
H^{R/L}  \\
P^{R/L}   \\
\phi^{R/L} \\
\Phi^{R/L} 
\end{matrix})
\ea
where 
\be
M^{R/L}
= \qty(\begin{matrix}
\mp\frac{4  s}{m^2}|\vec{k}|^3 & 1 - \frac{4 s^2 }{m^2}|\vec{k}|^2  &  \pm c\qty(2 |\vec{k}| - \frac{8 s^2}{m^2}|\vec{k}|^3)  & -\frac{4 cs}{m^2}|\vec{k}|^2 \\
	- |\vec{k}|^2 +\frac{4 s^2 }{m^2}|\vec{k}|^4 & \pm\frac{4  s}{m^2}|\vec{k}|^3 & 
 \frac{8 cs}{m^2}|\vec{k}|^4 & \pm\frac{4  c}{m^2}|\vec{k}|^3 \\
	\mp\frac{4  c}{m^2}|\vec{k}|^3 & -\frac{4  cs}{m^2}|\vec{k}|^2 & \mp\frac{8  c^2s}{m^2}|\vec{k}|^3 &-1 -\frac{4  c^2}{m^2} |\vec{k}|^2  \\
	\frac{8  cs}{m^2}|\vec{k}|^4 
 & \mp c\qty(2 |\vec{k}| - \frac{8 s^2}{m^2}|\vec{k}|^3) 	 & m^2- (3+4s^2)|\vec{k}|^2 +      \frac{16 s^2c^2}{m^2}|\vec{k}|^4 
 &\pm \frac{8  c^2s}{m^2}|\vec{k}|^3  \\
\end{matrix})
\,.
\ee
Here we have abbreviated $c = \cosh(2\xi)$ and $s = \sinh(2\xi)$. 

The solution of the system is given by $X^{R/L}=\exp(M^{R/L} \tau) X^{R/L}_0$, where $\tau$ denotes the time parameter, and $X^{R/L}=(H^{R/L},P^{R/L},\phi^{R/L},\Phi^{R/L})$, with $X^{R/L}_0$ denoting the initial values.

The matrices $M^{R/L}$  are diagonalizable and for both the $R$- and $L$-polarization the eigenvalues are given by
\be
\left(-\imath|\vec{k}|,+\imath|\vec{k}|,-\sqrt{-|\vec{k}|^2 - m^2},+\sqrt{-|\vec{k}|^2 - m^2}\right) \quad .
\ee
Remember that the eigenvalues for a harmonic oscillator with frequency $\omega$ are given as $(+\imath \omega, -\imath \omega)$. We therefore see that we have two massless propagating modes and two massive propagating modes, and that the dynamics of the transverse-traceless modes is stable.

 It is astonishing that, despite the coupling between the length metric and the massive $\phi_{ab}$ field with the ``wrong sign" for the kinetic and mass term, we obtain the same spectrum as for the non-coupled system. We thus also find that the eigenvalues are independent of the coupling parameter $\xi$.

The eigenvectors can be also computed explicitly. Note that the matrices $M^{R/L}$ are real. Therefore we can choose the eigenvectors associated to complex conjugates eigenvalues, to be also complex conjugated to each other. Here we only  need to consider their expansion in $1/m$. The eigenvectors for the eigenvalue $-\imath |\vec{k}|$ are given by
\ba
V^{R/L}_1&=&\left(1,\,\pm \imath |\vec{k}| \pm 4 (-\imath +s) |\vec{k}|^3 \frac{1}{m^2} ,\,
2\imath c |\vec{k}|^2 \frac{1}{m^2},
\mp 2 c (1-2\imath s)|\vec{k}|^3 \frac{1}{m^2}
\right) +{\cal O}\left(\frac{1}{m^3}\right)\, ,
\ea
whereas the eigenvector for the eigenvalue $+\imath |\vec{k}|$ is given by $V_2^{R/L}=\overline{V^{R/L}_1}$.
 We see that for $m\rightarrow \infty$ (that is for $m^2\gg k^2$), we  approach the dynamics of the pure gravitational system~\footnote{Remember, however, that $H_{ab}$ involves the $h$- and $\chi$-fields and $P_{ab}$ involves (time or spatial) derivatives of the $h$-,$\chi$- and $\phi$-fields.}, given by a massless degree of freedom in the $R$ and $L$ polarization respectively.

The eigenvectors for the eigenvalue $-\sqrt{-|\vec{k}|^2 - m^2}$ are given by
\ba
V^{R/L}_3&=&\left(
\pm 2|\vec{k}| c \frac{1}{m^2},0 , \frac{-\imath}{m},1\right)+{\cal O}\left(\frac{1}{m^3}\right)
\ea
and for the eigenvalue $\sqrt{-|\vec{k}|^2 - m^2}$ we have $V_4^{R/L}=\overline{V^{R/L}_3}$. 
For $m\rightarrow \infty$ we obtain an oscillator with mass $m$, in the $R$ and $L$ polarization respectively. To higher order in $1/m$ the eigenmode has a non-vanishing $H^{R/L}$ component. 

~\\
We have also a closed system of equations of motions for the $(l\pm)$-modes, modulo a term in the shift parameters $h_{0a}$. This term reflects that we have first-class constraints, which imply
\ba
\hat k^\theta_bC^b=0 \quad \text{and}\quad  k^\varphi_bC^b=0
\quad\quad\Rightarrow \quad\quad P^{l+}=0 \quad \text{and}\quad  P^{l-}=0 \quad.
\ea
One can furthermore use the mode expansion for the equation determining $\dot P_{ab}$ in (\ref{eq:HamiltonianEOM}) to show that $\dot P^{l+}=0$ and $\dot P^{l-}=0$. (This also follows from (\ref{commC}).)  

The remaining equations split into two parts.
Firstly we have that 
\ba\label{lpDynamics}
\qty(\begin{matrix}
		\dot{\phi}^{l\pm} \\
				\dot{\Phi}^{l\pm} \\
\end{matrix}) 
&=&
\qty(\begin{matrix}
\mp\frac{ c^2 s }{m^2}|\vec{k}|^3 &  -1-\frac{c^2}{m^2}|\vec{k}|^2 \\
m^2-s^2 |\vec{k}|^2+\frac{c^2s^2}{m^2}|\vec{k}|^4&\pm\frac{ c^2 s}{m^2} |\vec{k}|^3 \\
\end{matrix})
 \cdot\,
\qty(\begin{matrix}
\phi^{l\pm} \\
\Phi^{l\pm} \\
\end{matrix})\quad .\quad
\ea

Secondly we have equations determining $\dot H^{l\pm}$, which also involve the shift parameters 
\ba
		\dot{H}^{l\pm} 
   &=& 
 \pm c|\vec{k}|(1-\frac{s}{m^2} |\vec{k}|^2 )\phi^{l\pm}   -\frac{cs}{m^2} |\vec{k}|^2  \Phi^{l\pm}
  \,\, +\,\, 2 \imath  
		k^c {h^d}_0 t^{l\pm}_{cd}  \quad. 
\ea

The matrix in (\ref{lpDynamics}) can again be diagonalized and the eigenvalues (for the $l+$ and $l-$ modes) are given by
\ba
\left( -\sqrt{-|\vec{k}|^2 - m^2},+\sqrt{-|\vec{k}|^2 - m^2}\right) \quad .
\ea
Thus we have again a stable dynamics, describing two propagating degrees of freedom with mass $m$. These oscillations for the $(l\pm)$-modes of $\phi$ and $\Phi$ induce a time evolution for $H^{l\pm}$. For this time evolution we have also gauge parameters ${h^d}_0$ appearing. These (and the initial data) can be chosen such that $H^{l\pm}=0$ throughout.

The eigenvectors of the matrix appearing in (\ref{lpDynamics}) can also be computed, their expansion in $1/m$ is as follows
\ba
V^{l\pm}_1=\left(-\frac{\imath}{m},1\right)+{\cal O}\left(\frac{1}{m^3}\right), \quad\quad
V^{l\pm}_2=\left(+\frac{\imath}{m},1\right)+{\cal O}\left(\frac{1}{m^3}\right) \quad .
\ea
In the $m\rightarrow\infty$ limit we indeed obtain the dynamics of free oscillators with mass $m$.

~\\
Finally, we are left with the $(ll)$- and $(tr)$-modes. The constraints $C$ and $C_b$ imply
\ba
C=0\quad\Rightarrow H^{ll}-\sqrt{2}H^{tr}=0 \quad\quad \text{and}\quad \quad C_bk^b=0 \quad \Rightarrow \sqrt{2}P^{ll}+P^{tr}=0 \quad .
\ea
Due to (\ref{commC}) we also know that the time derivative of these constraints vanishes. It is therefore sufficient to consider the time evolution of the $(ll)$-mode for the $H$- and $P$-fields. The same holds for the $\phi$- and $\Phi$-fields as the $(tr)$-mode vanishes for these fields. 

Here we find a further decoupling of the equations of motions. The time evolution of the $H$- and $P$-fields is described by 
\ba
\dot{H}^{ll}=P^{ll}+2\imath \sqrt{\tfrac{2}{3}}k^d h_{d0} \,, \quad \quad
\dot{P}^{ll}= \sqrt{\tfrac{2}{3}}k^2 h_{00} \quad .
\ea
We can again choose initial data and gauge parameters such that $H^{ll}=0$ and $P^{ll}=0$ (and thus $H^{tr}=0$ and $P^{tr}=0$). 

The time evolution of the $\phi$- and $\Phi$-field is described by
\ba
\dot{\phi}^{ll}=-\Phi^{ll} \, ,\quad\quad
\dot{\Phi}^{ll}= (k^2+m^2)\phi^{ll} \quad .
\ea
This again describes a stable propagating degree of freedom with mass $m^2$.

In summary, we find that all propagating degrees of freedom show a stable oscillating dynamics. The eigenfrequencies are the same as for the non-coupled system, and hence do not depend on the coupling parameter $\xi$.  This explains why we found in Section \ref{Sec:EffectiveLengthMetricActions}, that integrating out the $\chi$- and $\phi$-fields one still just finds two massless propagating modes without ghosts.

 The Hamiltonian (\ref{Hprime}) indeed appears already in the form of non-coupled oscillators if we consider the limit $m\rightarrow \infty$.  The harmonic oscillator Hamiltonian for the $(\phi,\Phi)$-fields appears however with a negative overall sign.

The resulting simple eigenfrequencies for finite mass mean that there exist coordinate transformations so that the new Hamiltonian still describes non-coupled harmonic oscillators, even for finite mass. The part of this new Hamiltonian, which describes the massive fields however comes also with a negative overall sign, whereas the transverse-traceless modes of the massless fields come with a positive overall sign.

We thus obtain a stable dynamics at quadratic order for the Lagrangian. But going to higher order one has to expect instabilities that arise from the higher order coupling between these modes with positive and negative energy. There might exist  fine-tuned choices for the higher-order terms in the Lagrangian that lead to a stable dynamics, but this has still to be investigated.

Here we chose to consider the dynamics using a mode decomposition and the (partially complex) $(R,L,l+,l-,ll,tr)$ basis. With this choice we have a decoupling between the $R$ and $L$ modes, and a decoupling between the $l+$ and $l-$ modes.  An alternative choice would be to use the real basis $(+,\cross, l\theta, l\varphi, ll,tr)$ basis. One of course finds again the same eigenfrequencies. But one has also a coupling between the $+$ and $\cross$ modes (and the $l\theta$ and $l\varphi$ modes), see Appendix \ref{AppA}.   Such a coupling does not appear in pure gravity, and could therefore lead to an observational signature for area metrics.

  \section{Discussion}\label{Sec:Discussion}

Area metrics have appeared in a number of approaches to quantum gravity. Here we constructed the space of possible  actions for cyclic area metrics, to quadratic order in the area-metric perturbations. We imposed (linearized) diffeomorphism invariance and allowed mass terms for the non-length degrees of freedom.  The cyclic area metric perturbations can be conveniently parametrized by a length-metric perturbation, as well as two times fives degrees of freedom which parametrize the trace-free selfdual and anti-selfdual parts of the non-length perturbations.

Taking into account the freedom of rescaling the length-metric and non-length fields, we found a four-dimensional space of the quadratic area-metric actions. Two of these four couplings are the masses for the selfdual and anti-selfdual parts, the other two couplings describe how strongly the (anti-)selfdual parts are coupled to the transverse traceless part of the length-metric fluctuations.  Restricting to parity-invariant actions reduces the available freedome to two couplings.  

Our derivation of the second order action is more transparent than in the (covariant) constructive gravity approach \cite{ConGrav2}. This is due to two ingredients: firstly the parametrization of the area metric perturbations in Sec.~\ref{Sec:Param} into different parts according to their $SO(4)$ transformation behavior. Secondly the usage of $SO(4)$ representation theory in Sec.~\ref{Sec:KinTerms}, which allows us to quickly identify all possible terms in the quadratic action. This avoids the need to use computer algebra systems or dedicated software packages as in \cite{ConGrav2}, and leads to a more transparent representation of the dynamics, which made the subsequent analysis and discussions possible.

A special subclass of theories identified in our work features a degenerate kinetic term and leads to a ghost-free propagator for the gravitons in the effective length metric theory. The same type of action was found to appear in the perturbative continuum limit of (effective) spin foams \cite{AR2} and from modified Plebanski theory \cite{BorDitt}. This special subset is described by two couplings: one coupling describes the strength of parity violations, the other coupling corresponds to the mass of the selfdual and anti-selfdual parts of the area metric perturbations.  The former coupling corresponds to the Barbero-Immirzi parameter in loop quantum gravity \cite{BIParameter}.  In the perturbative continuum limit of spin foams, the mass is of the order of the Planck mass and may also depend on the Barbero-Immirzi parameter.

{The appearance of the Barbero-Immirzi parameter in loop quantum gravity is somewhat surprising as it parametrizes a family of canonical transformations in the first-order formulation of general relativity. These transformations can however not be implemented unitarily in loop quantum gravity \cite{Rovelli:1997na}. But the appearance of the Barbero-Immirzi parameter can also be understood as a result of the extension of the quantum configuration space from length metrics to area metrics in loop quantum gravity \cite{Ryan3,EffSF1,Padua23}. In this paper we found that the Barbero-Immirzi parameter can indeed be identified as one of the couplings appearing in area-metric gravity theories, where it already affects the classical dynamics.

These results hold for both Euclidean and Lorentzian signature. Crucially, we find however that the Lorentzian, but not the Euclidean, area-metric action has indefinite kinetic and mass terms for the degrees of freedom encoding the self-dual and anti-selfdual parts of the area-metric perturbations.  As these degrees of freedom couple to the length metric, one might expect an unstable dynamics even for the free theory. This seems however to contradict the finding of a ghost-free propagator for the effective length-metric action. We therefore performed a canonical analysis for the actions leading to a ghost-free propagator. Here one finds that the degeneracy of the kinetic  term, together with having mass terms, leads to second-class constraints which remove half of the selfdual and anti-selfdual degrees of freedom.  The canonical analysis confirms, however, that the remaining part propagates and that it has negative definite kinetic and mass terms. One nevertheless obtains a stable dynamics, which moreover separates into two massless eigenmodes (with positive energy) and five massive eigenmodes (with negative energy).  One can also find a mixing between the cross- and plus-polarizations of the gravitons in area-metric dynamics, which is parametrized by the Barbero-Immirzi parameter. This does not appear in general relativity, and might therefore lead to an observational signature for area metrics and allow for constraining the Barbero-Immirzi parameter.

A crucial question in determining whether area-metric theories with real~\footnote{The (effective) spin foam action  includes imaginary terms in Euclidean \cite{EffSF1} and Lorentzian \cite{EffSF3} signature.} actions are viable in Lorentzian signature will be to see whether this stability can be extended to higher order in the perturbative expansion of area metrics, or whether this stability holds also for non-flat backgrounds.  For backgrounds admitting a Wick rotation one can connect Euclidean to Lorentzian solutions. One could therefore hope to find stability for such backgrounds. 

A similar change~\footnote{We thank Laurent Freidel for pointing this out.} in definiteness appears for the Kodama state, which is conjectured to describe a vacuum for quantum gravity \cite{Kodama}. This state is not normalizable (in Fock-space) in Lorentzian signature. The root for this non-normalizabilty is that this state prescribes negative-helicity gravitons to have negative energy, and  positive-helicity gravitons to have positive energy \cite{WittenKodama}. On the other hand, the Euclidean Kodama state is delta-function normalizable \cite{FreiSmoKodama}.  The question of whether one can find a non-perturbative inner product in which the Lorentzian Kodama state is normalizable, is still open \cite{FreiAlex23}.

\begin{acknowledgments}

We thank Luca Buoninfante, Laurent Freidel, Aaron Held and Simone Speziale for enlightening discussions. JNB is supported by an NSERC grant awarded to BD.  KK is grateful to the Perimeter Institute for hospitality during the period when this collaboration was initiated. Research at Perimeter Institute is supported in part by the Government of Canada through the Department of Innovation, Science and Economic Development Canada and by the Province of Ontario through the Ministry of Colleges and Universities.

\end{acknowledgments}

\begin{appendix}

\section{A different choice of polarization basis}\label{AppA}

In Section~\ref{Sec:SolH} we solved the Hamiltonian equations of motions, using a mode expansion.
Here we will use an alternative basis of modes for the transverse-traceless sector, given by the plus and cross polarization tensors $t^+_{ab}$ and $t^{\cross}_{ab}$, see equation~(\ref{PlusCross}).

Different from the case with right and left polarization, the dynamics couples the plus- and cross-modes. The time evolution is described by
\ba
\qty(\begin{matrix}
	\dot{H}^+  \\
	\dot{H}^{\cross}\\
	\dot{P}^+   \\
	\dot{P}^{\cross} \\
		\dot{\phi}^+ \\
			\dot{\phi}^{\cross} \\
				\dot{\Phi}^{+} \\
				\dot{\Phi}^{\cross} 
\end{matrix}) 
&=&
M^{TT} \, \cdot\,
\qty(\begin{matrix}
H^+  \\
H^{\cross}\\
P^+   \\
P^{\cross} \\
\phi^+ \\
\phi^{\cross} \\
\Phi^{+} \\
\Phi^{\cross} 
\end{matrix})
\ea
where $M^{TT}=(M_1 \, M_2)$ with
\be
M_1 
= \qty(\begin{matrix}
	0 & \frac{4 \imath s}{m^2}|\vec{k}|^3 & 1 - \frac{4 s^2 }{m^2}|\vec{k}|^2 & 0 \\
	-\frac{4 \imath s}{m^2}|\vec{k}|^3 & 0 & 0 & 1 - \frac{4 s^2}{m^2}|\vec{k}|^2  \\
	-|\vec{k}|^2 + \frac{4}{m^2}|\vec{k}|^4	 & 0 & 0 & -\frac{4 \imath s}{m^2} |\vec{k}|^3  \\
	0 & -|\vec{k}|^2 + \frac{4}{m^2}|\vec{k}|^4	 & \frac{4 \imath s}{m^2}|\vec{k}|^3 & 0 \\
	0 & \frac{4 \imath c}{m^2}|\vec{k}|^3 & -\frac{4 cs}{m^2}|\vec{k}|^2 & 0  \\
	-\frac{4 \imath c }{m^2}|\vec{k}|^3 & 0 & 0 & -\frac{4 cs}{m^2}|\vec{k}|^2  \\
	\frac{8 cs}{m^2}|\vec{k}|^4 & 0 & 0 & c\qty(2 \imath |\vec{k}| - \frac{8 \imath s^2}{m^2}|\vec{k}|^3)  \\
	0 & \frac{8 c s|\vec{k}|^4}{m^2} & c\qty(-2 \imath |\vec{k}| +\frac{8 \imath s^2}{m^2}|\vec{k}|^3) & 0  \\
\end{matrix})
\,.
\ee
and
\be
M_2&=&
\qty(\begin{matrix}
 0& c\qty(-2 \imath |\vec{k}| + \frac{8 \imath s^2}{m^2} |\vec{k}|^3) & -\frac{4 c s}{m^2} |\vec{k}|^2 & 0 \\
   c\qty(2 \imath |\vec{k}| - \frac{8 \imath s^2}{m^2} |\vec{k}|^3)  & 0 & 0 & -\frac{4 cs}{m^2}{|\vec{k}|^2}\\
    \frac{8 c s}{m^2}|\vec{k}|^4 & 0 & 0 & -\frac{4 \imath c}{m^2}|\vec{k}|^3\\
     0 & \frac{8 cs}{m^2}|\vec{k}|^4 & \frac{4 \imath c}{m^2}|\vec{k}|^3  & 0 \\
      0 & \frac{8 \imath c^2 s}{m^2} |\vec{k}|^3 & -1 - \frac{4  c^2}{m^2}  |\vec{k}|^2 & 0\\
       -\frac{8 \imath c^2 s}{m^2}|\vec{k}|^3 & 0 & 0 & -1 - \frac{4 c^2 }{m^2}  |\vec{k}|^2\\
        \qty(1-4c^4)|\vec{k}|^2 + m^2 \frac{16 c^2 s^2}{m^2}|\vec{k}|^4 & 0 & 0 & -\frac{8 \imath c^2 s}{m^2}|\vec{k}|^3\\
         0 & \qty(1-4c^2)|\vec{k}|^2 + m^2 +\frac{16 c^2 s^2}{m^2}|\vec{k}|^3 & \frac{8 \imath c^2 s }{m^2}|\vec{k}|^3  & 0\\
\end{matrix}) \, .
\ee
As before,  we have abbreviated $c = \cosh(2\xi)$ and $s = \sinh(2\xi)$. 
The eigenvalues of $M^{TT}$ coincide with the ones found in Section~\ref{Sec:SolH}
\be
\left(-\imath|\vec{k}|,+\imath|\vec{k}|,-\imath|\vec{k}|,+\imath|\vec{k}|,-\sqrt{-|\vec{k}|^2 - m^2},+\sqrt{-|\vec{k}|^2 - m^2},-\sqrt{-|\vec{k}|^2 - m^2},+\sqrt{-|\vec{k}|^2 - m^2}\right) \quad .
\ee
The first pair of eigenvalues comes with the following eigenvectors
\ba
V_1&=&\left(1,\,0,\,-\imath |\vec{k}| + 4\imath |\vec{k}|^3\frac{1}{m^2} ,\,+4\imath s|\vec{k}|^3\frac{1}{m^2} , \,0, \,+2 c |\vec{k}|^2 \frac{1}{m^2},\,+4\imath s c |\vec{k}|^3\frac{1}{m^2},\,-2\imath c|\vec{k}|^3\frac{1}{m^2}\right) +{\cal O}\left(\frac{1}{m^3}\right)\, , \nn\\
V_2&=&\left(1,\,0,\,+\imath |\vec{k}| - 4\imath |\vec{k}|^3\frac{1}{m^2} ,\,+4\imath s|\vec{k}|^3\frac{1}{m^2} , \,0, \,-2 c|\vec{k}|^2 \frac{1}{m^2},\,-4\imath s c |\vec{k}|^3\frac{1}{m^2},\,-2\imath c|\vec{k}|^3\frac{1}{m^2}\right) +{\cal O}\left(\frac{1}{m^3}\right) ,
\ea
and the next pair of eigenvalues is associated to
\ba
V_3&=&\left(
0,\,1,\, -4\imath s |\vec{k}|^3 \frac{1}{m^2} ,\, -\imath |\vec{k}| + 4\imath |\vec{k}|^3\frac{1}{m^2},\, -2|\vec{k}|^2c \frac{1}{m^2},0, +2\imath c |\vec{k}|^3\frac{1}{m^2},+4\imath s c|\vec{k}|^3\frac{1}{m^2}
\right) +{\cal O}\left(\frac{1}{m^3}\right) \, ,\nn\\
V_4&=&\left(
0,\,1,\, -4\imath s |\vec{k}|^3 \frac{1}{m^2} ,\, +\imath |\vec{k}| - 4\imath |\vec{k}|^3\frac{1}{m^2},\, +2|\vec{k}|^2c \frac{1}{m^2},0, +2\imath c |\vec{k}|^3\frac{1}{m^2},-4\imath s c|\vec{k}|^3\frac{1}{m^2}
\right) +{\cal O}\left(\frac{1}{m^3}\right) \,.\quad\quad 
\ea
For $m\rightarrow \infty$ (that is for $m^2 \gg k^2$), we approach the same dynamics of for the pure gravitational system~\footnote{Remember however that $H_{ab}$ does involve the $h$- and $\chi$-fields and $P_{ab}$ involves (time or spatial) derivatives of the $h$-,$\chi$- and $\phi$-fields.}, given by a massless degree of freedom in the plus and cross polarization respectively.

Going to higher order in the  $1/m$ expansion, we can also observe that the presence of the area metric degrees of freedom induces a mixing between the plus and cross polarizations (for $s=\sinh(2\xi)\neq0$) $H^+$ and $H^{\cross}$.

 The next four eigenvectors are given as
 \ba
V_5&=&\left(
0,\, -2 c |\vec{k}|\frac{1}{m^2},\, 0,\,0\,,\frac{1}{m},\,0, +\imath -\frac{\imath}{2}(7 |\vec{k}|^2 +8 s^2 |\vec{k}|^2)\frac{1}{m^2},\,0
\right)+{\cal O}\left(\frac{1}{m^3}\right)\, ,
\nn\\
V_6&=&\left(
0,\, +2 c |\vec{k}|\frac{1}{m^2},\, 0,\,0\,,\frac{1}{m},\,0, -\imath +\frac{\imath}{2}(7 |\vec{k}|^2 +8 s^2 |\vec{k}|^2)\frac{1}{m^2},\,0
\right)+{\cal O}\left(\frac{1}{m^3}\right)
 \ea
and
\ba
V_7&=&
\left(
+2 c |\vec{k}|\frac{1}{m^2},\,0,\,0,\,0,\,0,\frac{1}{m},\,0,\, +\imath -\frac{\imath}{2}(7 |\vec{k}|^2 +8 s^2 |\vec{k}|^2)\frac{1}{m^2} 
\right) +{\cal O}\left(\frac{1}{m^3}\right)\, ,
\nn\\
V_8&=&
\left(
-2 c |\vec{k}|\frac{1}{m^2},\,0,\,0,\,0,\,0,\frac{1}{m},\,0,\, -\imath +\frac{\imath}{2}(7 |\vec{k}|^2 +8 s^2 |\vec{k}|^2)\frac{1}{m^2}  
\right) +{\cal O}\left(\frac{1}{m^3}\right) \, .
\ea
For $m\rightarrow \infty$ we obtain an oscillator with mass $m$, in the plus and cross polarization respectively.  To higher order in $1/m$ the  $\phi^+$ dominated eigenmodes have a non-vanishing $H^{\cross}$ component, and the 
$\phi^{\cross}$ dominated eigenmodes have a non-vanishing $H^{+}$ component

Considering only the dynamics of the massless modes, we noticed that these feature a mixing of the plus and cross polarization, which in this form does not appear in linearized general relativity and could lead to a possible observational signature of area metric dynamics.

\end{appendix}

\end{document}